\documentclass[twocolumn,showpacs,amsmath,amssymb,superscriptaddress,nofootinbib]{revtex4-1}
 
\pdfoutput=1

\usepackage{graphicx}
\usepackage{dcolumn}
\usepackage{amssymb}
\usepackage{pifont}

\usepackage{bm}
\usepackage[usenames,dvipsnames]{color}
\usepackage[colorlinks=true, linkcolor=BrickRed, citecolor=Blue, urlcolor=Blue, filecolor=Blue]{hyperref}
\usepackage[dvipsnames]{xcolor}

\def\nn{\nonumber}
\newcommand{\be}{\begin{equation}}
\newcommand{\ee}{\end{equation}}
\newcommand{\beq}{\begin{equation}}
\newcommand{\eeq}{\end{equation}}
\newcommand{\bea}{\begin{eqnarray}}
\newcommand{\eea}{\end{eqnarray}}
\newcommand{\beaa}{\begin{eqnarray*}}
\newcommand{\eeaa}{\end{eqnarray*}}
\newcommand{\ba}{\begin{array}}
\newcommand{\ea}{\end{array}}
\newcommand{\bi}{\begin{itemize}}
\newcommand{\ei}{\end{itemize}}
\newcommand{\ben}{\begin{enumerate}}
\newcommand{\een}{\end{enumerate}}

\newcommand{\p}{\partial}
\newcommand{\ds}{{\sf DarkSUSY}}

\begin{document}

\title{Early kinetic decoupling of dark matter:\\ 
when the standard way of calculating the  thermal relic density fails}

\author{Tobias Binder}
\email{tobias.binder@ipmu.jp}
\affiliation{Institute for Theoretical Physics, Georg-August University G\"ottingen, Friedrich-Hund-Platz 1, D-37077 G\"ottingen, Germany
}

\author{Torsten Bringmann}
\email{torsten.bringmann@fys.uio.no}
\affiliation{Department of Physics, University of Oslo, Box 1048, NO-0316 Oslo, Norway}

\author{Michael Gustafsson}
\email{michael.gustafsson@theorie.physik.uni-goettingen.de}
\affiliation{Institute for Theoretical Physics, Georg-August University G\"ottingen, Friedrich-Hund-Platz 1, D-37077 G\"ottingen, Germany
}
\author{Andrzej Hryczuk}
\email{a.j.hryczuk@fys.uio.no}
\affiliation{Department of Physics, University of Oslo, Box 1048, NO-0316 Oslo, Norway}

\date{May 8, 2020}

\begin{abstract} 

Calculating the abundance of thermally produced dark matter particles has become a standard procedure, with 
sophisticated methods guaranteeing a precision that matches the percent-level accuracy in the observational 
determination of the dark matter density. 
Here, we point out that one of the main assumptions in the commonly
adopted formalism, namely local thermal equilibrium during the freeze-out of annihilating dark matter
particles, does not have to be satisfied in general. We present two methods for how to deal with such 
situations, in which the kinetic decoupling of dark matter happens so early that it interferes with the
chemical decoupling process: {\it i)} an approximate treatment in terms of a coupled system of differential 
equations for the leading momentum moments of the dark matter distribution, and {\it ii)} a full numerical solution of
the Boltzmann equation in phase-space. For illustration, we apply these methods to the case of 
Scalar Singlet dark matter. We explicitly show that even in this simple model
the prediction for the dark matter abundance can be affected by up to one order of magnitude
compared to the traditional treatment.

\end{abstract}


\maketitle

\section{Introduction}

The leading hypothesis for the cosmological dark matter (DM) \cite{Hinshaw:2012aka,Ade:2015xua} 
is a new type of elementary particle \cite{Bertone:2004pz}. One of the most attractive 
options to explain the present abundance of these particles consists in the 
possibility that they have been thermally produced in the early universe. This is
particularly interesting for the scenario originally studied by Lee and 
Weinberg \cite{Lee:1977ua}, as well as 
others \cite{Hut:1977zn,Sato:1977ye,Dicus:1977nn,Wolfram:1978gp}, 
in which non-relativistic DM
particles initially are kept in thermal equilibrium with the heat bath through frequent 
annihilation and creation processes with standard model (SM) particles. Once the 
interaction rate starts to fall behind the expansion rate of the universe, 
the DM number density begins to `freeze out' and remains covariantly conserved. For 
weakly interacting massive particles (WIMPs), elementary particles with masses 
and interaction strengths at the electroweak
scale, this scenario automatically leads to a relic abundance in rough
agreement with the observed DM density  --- a fact sometimes
referred to as the {\it WIMP miracle}.

The by now standard treatment \cite{Gondolo:1990dk,Edsjo:1997bg} of calculating the 
resulting DM abundance in these scenarios implements an efficient and highly accurate 
method of solving the Boltzmann equation 
for a given (effective) invariant DM annihilation rate.
This approach fully captures, in particular, the three famous exceptions to the original relic density 
calculations pointed out in a seminal paper by Griest and Seckel \cite{Griest:1990kh},
namely co-annihilations, threshold effects and resonances. The main assumption 
entering this formalism is that, during the freeze-out process, DM is still kept in {\it local} 
thermal equilibrium with the heat bath by frequent scattering processes with relativistic 
SM particles. For many WIMP candidates, this is indeed satisfied to a high accuracy
and kinetic decoupling typically only happens much later than the chemical decoupling
\cite{Bringmann:2009vf}.

\smallskip

Here we point out that exceptions to this standard lore do exist, even in very simple 
scenarios, where kinetic decoupling happens so early that it cannot be neglected 
during the freeze-out process. We develop both semi-analytical and fully numerical 
methods to solve the Boltzmann equation and to compute the DM relic abundance in 
these cases. Technically, one of the challenges that had to be overcome for obtaining
sufficiently accurate results  was to extend the highly non-relativistic Boltzmann equation, 
as discussed previously in the literature, to the semi-relativistic regime.
Numerically, we also succeeded to resolve the evolution of the full phase-space distribution
accurately enough to test, for the first time, the underlying assumptions for the standard
way of calculating the relic density of WIMPs or other self-annihilating DM candidates
(for a recent example where the relic density is instead set by inelastic scattering, rather than
self-annihilation, see Ref.~\cite{DAgnolo:2017dbv}). 
We illustrate our general results by a detailed discussion of the 
Scalar Singlet model \cite{Silveira:1985rk,McDonald:1993ex,Burgess:2000yq}, for which 
we find a DM relic density that differs by up to an order of magnitude from the standard 
treatment.

\smallskip

This article is organized as follows. In Section  \ref{sec:BEgeneral},
we start with a general description of the underlying Boltzmann equation that governs 
the DM phase-space evolution. We then briefly review the standard treatment of 
solving for the DM number density
(Section \ref{sec:standard}), extend this by deriving a {\it coupled} system of evolution 
equations for the number density and the velocity dispersion (Section \ref{sec:cBE}),
and finally introduce our framework for a fully numerical solution (Section \ref{sec:magicmichael}).
Section \ref{sec:singlet} is devoted to a thorough application of these methods to the Scalar Singlet
model. 
We comment on our results in Section \ref{sec:disc},
and discuss potential other areas of application, before we conclude in Section \ref{sec:conc}.
In two  Appendices we discuss in 
detail the evolution of the Singlet DM phase-space 
density for selected parameter points (App.~\ref{app:SingletDetails}) and comment on the semi-relativistic 
form of the scattering operator in the Boltzmann equation (App.~\ref{app:semirel}).

\section{Thermal production of dark matter}
\label{sec:BEgeneral}

Let us denote the DM particle by $\chi$, and its phase-space density by
$f_\chi(t,\mathbf{p})$. The evolution of $f_\chi$ is governed by the Boltzmann
equation which, in an expanding Friedmann-Robertson-Walker universe,
is given by \cite{Kolb:1990vq,Bringmann:2006mu}
\be
  \label{diff_boltzmann}
  E\left(\partial_t-H\mathbf{p}\cdot\nabla_\mathbf{p}\right)f_\chi=C[f_\chi]\,.
\ee
Here, $H=\dot a/a$ is the Hubble parameter, $a$ the scale factor, and the collision term $C[f_\chi]$
contains all interactions between DM and SM particles $f$. For
WIMPs, we are to leading order interested in two-body processes for
DM annihilation and elastic scattering, $C=C_{\rm ann}+C_{\rm el}$, where
\bea
  \label{Candd_def}
  C_\mathrm{ann}&=&\frac{1}{2g_\chi}\int\frac{d^3\tilde p}{(2\pi)^32\tilde E}\int\frac{d^3k}{(2\pi)^32\omega}\int\frac{d^3\tilde k}{(2\pi)^32\tilde \omega}\\
&&\times(2\pi)^4\delta^{(4)}(\tilde p+p-\tilde k-k)\nonumber\\
&&\times\left[
\left|\mathcal{M}\right|^2_{\bar\chi\chi\leftarrow \bar f f}g(\omega)g(\tilde \omega)
-\left|\mathcal{M}\right|^2_{\bar\chi\chi\rightarrow \bar f f}f_\chi(E)f_\chi(\tilde E)
\right]\,,\nonumber
\eea
and
\bea
  \label{Celd_ef}
  C_\mathrm{el}&=&\frac{1}{2g_\chi}\int\frac{d^3k}{(2\pi)^32\omega}\int\frac{d^3\tilde k}{(2\pi)^32\tilde \omega}\int\frac{d^3\tilde p}{(2\pi)^32\tilde E}\\
  &&\times(2\pi)^4\delta^{(4)}(\tilde p+\tilde k-p-k) {\left|\mathcal{M}\right|}^2_{\chi f\leftrightarrow\chi f}\nonumber\\
  &&\times\left[(1\mp g^\pm)(\omega)\, g^\pm(\tilde\omega)f_\chi(\mathbf{\tilde p})-
  (\omega\leftrightarrow\tilde\omega, \mathbf{p}\leftrightarrow\mathbf{\tilde p})\right]\,.\nonumber
\eea
In the above expressions, ${\left|\mathcal{M}\right|}^2$ refers to the respective squared 
amplitude, summed
over {\it all}  spin and other internal degrees of freedom, as well as all SM particles $f$. We assume the SM
particles to be in thermal equilibrium, 
such that their phase-space distribution is given
by $g^\pm(\omega)=1/\left[\exp(\omega/T)\pm1 \right]$. Note that we have neglected Bose enhancement 
and Pauli blocking factors for $f_\chi$ here, as we assume  DM to be nonrelativistic; momentum 
conservation then implies that,  in $C_\mathrm{ann}$, we can also neglect these factors for the SM particles.

Assuming $CP$ invariance, and using the fact that in thermal equilibrium annihilation and creation
processes should happen with the same frequency, the annihilation term given by Eq.~(\ref{Candd_def}) 
can be further simplified to \cite{Gondolo:1990dk}
\bea
  \label{Cann_simp}
  C_\mathrm{ann}&=& g_{\chi} E\int\frac{d^3\tilde p}{(2\pi)^3} \,v \sigma_{\bar\chi\chi\rightarrow \bar f f}\nonumber\\
  &&\times\left[
f_{\chi,{\rm eq}}(E)f_{\chi, {\rm eq}}(\tilde E)-f_\chi(E)f_\chi(\tilde E)
\right]\,, 
\eea
where $v=v_{\rm M\o l}\equiv ({E \tilde E})^{-1}[{(p \cdot \tilde p)^2-m_\chi^4}]^{1/2}$ is the M\o ller velocity,
which in the rest frame of one of the DM particles coincides with the lab velocity $v_{\rm lab}=[s(s-4m_\chi^2)]^{1/2}/{(s-2m_\chi^2)}$.

The scattering term, on the other hand, is in general considerably more difficult to manage. Analytic expressions have,
however,  been
obtained in the highly non-relativistic limit of the DM particles, and assuming that the momentum transfer in
 the scattering process is much smaller than the DM mass \cite{Bringmann:2006mu, Bringmann:2009vf, Kasahara:2009th, Gondolo:2012vh, Binder:2016pnr, Bringmann:2016ilk}:
 \be
 \label{Cresult}
C_\mathrm{el}\simeq 
\frac{m_{\chi}}{2} \gamma(T)
{\Bigg [}
T m_{\chi} \partial_p^2 + \left( p + 2 T  \frac{m_{\chi}}{p} \right) \partial_p + 3
{\Bigg ]}f_{\chi}\,,
\ee
where the momentum exchange rate is given by
\be
  \label{cTdef}
  \gamma(T) =  \frac{1}{48 \pi^3g_\chi m_\chi^3} \int d\omega\,g^\pm
  \partial_\omega\left( k^4 
  \left<\left|\mathcal{M}\right|^2\right>_t\right),
\ee
with
\be  
\left\langle\left|\mathcal{M}\right|^2\right\rangle_t 
\equiv \frac{1}{8k^4}\int_{-4{k}_\mathrm{cm}^2}^0
\!\!\!\! dt(-t)\left|\mathcal{M}\right|^2
= 16 \pi m_\chi^2\,\sigma_T\,,
\label{eq:sigmaT}
\ee
and ${k}_\mathrm{cm}^2  \!=\! \left(s-(m_\chi-m_f)^2\right) \left(s-(m_\chi+m_f)^2\right)/(4 s)$ evaluated at ${s=m_\chi^2+2\omega m_\chi+m_f^2}$. 
Here, $\sigma_T=\int d\Omega (1-\cos\theta)d\sigma/d\Omega$ is the standard {\it transfer cross section} for elastic scattering.
In Appendix \ref{app:semirel}, we discuss how the scattering term is expected to change
in the semi-relativistic case, i.e.~when the assumption of highly non-relativistic DM is slightly
relaxed. For reference, we will in the following use
\be
C_\mathrm{el}\simeq 
\frac{E}{2} \gamma(T)
{\Bigg [}
T E \partial_p^2 + \left( p + 2 T  \frac{E}{p} + T \frac{p}{E} \right) \partial_p + 3
{\Bigg ]}f_{\chi}
\label{Csemirel}
\ee
when explicitly addressing this regime. 

\subsection{The standard treatment}
\label{sec:standard}

In order to calculate the DM relic abundance, we can integrate the Boltzmann Eq.~(\ref{diff_boltzmann}) 
over $\mathbf{p}$. This results in
\be
\label{eq:boltzsimp}
\frac{dn_\chi}{dt}+3Hn_\chi= g_\chi \int \frac{d^3p}{(2\pi)^3E} C_{\rm ann}[f_\chi]\,,
\ee
which has to be solved for the DM number density 
\be
n_\chi=g_\chi\int d^3p/(2\pi)^3\,f_\chi(\mathbf{p})
\label{eq:n}
\ee
(note that $C_{\rm el}$ vanishes once it is integrated over).
In order to evaluate the r.h.s.~of this equation, the usual assumption \cite{Gondolo:1990dk}
is that during chemical freeze-out one can make the following ansatz for the DM distribution:
\be
\label{kineq_ansatz}
f_\chi=A(T)f_{\chi,{\rm eq}}=\frac{n_\chi}{n_{\chi, {\rm eq}}} f_{\chi,{\rm eq}}\,,
\ee
where $A(T)=1$ in full equilibrium, i.e.~before chemical freeze-out. 
This is motivated by the fact DM-SM scattering typically proceeds 
at a much faster rate than DM-DM annihilation, because the number density of relativistic SM particles
is not Boltzmann suppressed like that of the non-relativistic DM particles. In that case, DM particles 
are kept in {\it local} thermal equilibrium even when the annihilation rate starts to fall behind the Hubble 
expansion and chemical equilibrium can no longer be maintained.

Approximating furthermore $f_{\chi,{\rm eq}}(E)\simeq \exp(-E/T)$, i.e.~neglecting the impact of 
quantum statistics for non-relativistic particles, five of the six integrals in Eq.~(\ref{eq:boltzsimp}) 
can be performed analytically. This by now standard treatment, as established
by Gondolo \& Gelmini \cite{Gondolo:1990dk}, results in the often-quoted expression 
\be
\label{boltzn_standard}
\frac{dn_\chi}{dt}+3Hn_\chi= \langle \sigma v\rangle\left(n_{\chi,{\rm eq}}^2 -n_\chi^2\right)\,,
\ee
where $n_{\chi,{\rm eq}}= g_\chi m_\chi^2 T K_2(m_\chi/T)/(2\pi^2)$ and
\bea
\label{therm_av1}
 \left\langle \sigma v\right\rangle
 &\equiv&
\frac{g_\chi^2}{n_{\chi,{\rm eq}}^2}
\int \frac{d^3p}{(2\pi)^3} \frac{d^3\tilde p}{(2\pi)^3} 
 \sigma v_{\bar\chi\chi\rightarrow \bar f f} 
f_{\chi,{\rm eq}}(\mathbf{p}) f_{\chi,{\rm eq}}(\tilde{\mathbf{p}})
\nonumber\\
\\
 &=&  \int_1^\infty\!\!\! d\tilde s\, \sigma_{\bar\chi\chi\rightarrow \bar f f}v_{\rm lab}
 \frac{  2m_\chi \sqrt{\tilde s\!-\!1}(2\tilde s\!-\!1)K_1\!\!\left(\frac{2{\sqrt{\tilde s}} m_\chi}{T}\right)}
 {{T K_2}^2(m_\chi/T)}\,.\label{therm_av1eq} \nonumber\\
\eea
Here, $K_i$ are the modified Bessel functions of order $i$, and we have introduced
$\tilde s\equiv s/(4m_\chi^2)$. While there are various ways to state the final result for
$\left\langle \sigma v\right\rangle$, the form given above stresses that physically 
one should indeed think of this quantity as a thermal average of $\sigma v_{\rm lab}$ rather than any other
combination of cross section and velocity
(in the sense that we strictly have 
$\left\langle \sigma v\right\rangle=\sigma v_{\rm lab}$ for $\sigma v_{\rm lab}=$\,const;
for e.g.~$\sigma v_{\rm CMS}=$\,const, on the other hand, with $v_{\rm CMS}=2\sqrt{1-4m_\chi^2/s}$
being the {\it relative} velocity in the CMS frame,  
we instead have $\left\langle \sigma v\right\rangle\to \sigma v_{\rm CMS}$  only 
in the limit $T\to0$).

By introducing dimensionless variables 
\bea
x&\equiv& m_\chi/T\,,\\
Y&\equiv& n_\chi/s\,,\label{Ydef}
\eea
and assuming entropy conservation, finally, the above Boltzmann equation for the 
number density, Eq.~(\ref{boltzn_standard}), can be brought into an alternative form 
that is particularly suitable for numerical integration:
\be
\label{boltzYsimp}
\frac{Y'}{Y}=\frac{sY}{x\tilde H}
\left\langle \sigma v\right\rangle
\left[
\frac{Y_{\rm eq}^2}{Y^2}-1
\right]\,.
\ee
Here, $s=(2\pi^2/45) g^s_{\rm{eff}}T^3$ denotes the entropy density, $'\equiv d/dx$ and 
$\tilde H\equiv H/\left[1+ \tilde g(x)\right]$ where
\be
\label{gtildedef}
\tilde g \equiv \frac{1}{3} \frac{T}{g^s_{\rm\text{eff}}}\frac{ d g^s_{\rm\text{eff}}}{d T} \,.
\ee
The value of $Y$ today, $Y_0\equiv Y(x\to \infty)$, can then be related to the observed DM 
abundance by \cite{Gondolo:1990dk}
\be
\label{eq:oh2}
\Omega_\chi h^2= 2.755\times10^{10}\left(\frac{m_\chi}{100\,{\rm GeV}}\right) \left(\frac{T_{\rm CMB}}{2.726\,{\rm K}}\right)^3Y_0\,.
\ee
We note that
Eq.~(\ref{boltzYsimp}) is the basis for the implementation of relic density 
calculations in all major numerical codes \cite{Belanger:2001fz,Gondolo:2004sc,Belanger:2006is,
Arbey:2009gu,Belanger:2013oya,ds6,Workgroup:2017lvb}.

\subsection{Coupled Boltzmann equations}
\label{sec:cBE}

The main assumption that enters the standard treatment reviewed above is 
contained in Eq.~(\ref{kineq_ansatz}), i.e.~the requirement that during chemical freeze-out,
or in fact during any period when the comoving DM density changes, local thermal
equilibrium with the heat bath is maintained. If that assumption is not justified, one
has in principle to solve the full Boltzmann equation in phase space, Eq.~(\ref{diff_boltzmann}), 
numerically (see next subsection). 
As first pointed out in Ref.~\cite{vandenAarssen:2012ag}, however, it sometimes suffices
to take into account the second moment of Eq.~(\ref{diff_boltzmann}), instead of only the zeroth moment 
as in the previous subsection. This leads to a 
relatively simple {\it coupled} system of differential equations that generalizes Eq.~(\ref{boltzYsimp}).

The starting point is to define, in analogy to $Y$ for the zeroth moment of $f_\chi$, a 
dimensionless version of the {\it second} 
moment of $f_\chi$:
\be
\label{ydef}
 y\equiv \frac{m_\chi}{3 s^{2/3}}
\left\langle \frac{\mathbf{p}^2}{E} \right\rangle
=
\frac{m_\chi}{3 s^{2/3}}
\frac{g_\chi}{n_\chi}\int \frac{d^3p}{(2\pi)^3}\,\frac{\mathbf{p}^2}{E} f_\chi(\mathbf{p})\,.
\ee
For a thermal distribution, 
the DM particles thus have a {\it temperature}
\be
\label{tdef}
  T_\chi = y s^{2/3}/m_\chi\,.
\ee
We note that for non-thermal distributions we could still view this last equation as an alternative {\it definition} 
of the DM `temperature', or velocity dispersion, in terms of the second moment of $f_\chi$ as introduced above.
This allows, e.g., a convenient characterization 
of kinetic decoupling as the time when $T_\chi$ 
no longer equals $T$ but instead
starts to approach the asymptotic scaling of $T_\chi=T_{\rm kd} (a/a_{\rm eq})^{-2}$ for highly non-relativistic DM 
\cite{Bringmann:2006mu,Bringmann:2009vf}.

Integrating Eq.~(\ref{diff_boltzmann}) over $g_\chi\int d^3p/(2\pi)^3/E$ and $g_\chi\int d^3p/(2\pi)^3\mathbf{p}^2/E^2$,
respectively, we find 
\bea
\frac{Y'}{Y} &=& \frac{m_\chi}{x \tilde H}C_0\,, \label{Yfinal}\\
\frac{y'}{y} &=& \frac{m_\chi }{x \tilde H} C_2 - \frac{Y'}{Y} 
+\frac{H}{x\tilde H}
\frac{\langle p^4/E^3 \rangle}{3T_\chi}\,,\label{yfinal}
\eea
where
\be
\langle p^4/E^3 \rangle \equiv n_\chi^{-1}~ g_\chi \int \frac{d^3p}{(2\pi)^3}\,\frac{\mathbf{p}^4}{E^3} f_\chi(\mathbf{p})
\label{p4E3def}
\ee
and we introduced the moments of the collision term as
\bea
m_\chi n_\chi C_0&\equiv&  g_\chi \int \frac{d^3p}{(2\pi)^3E}\, C[f_\chi]\,,\\
m_\chi n_\chi \left\langle \frac{\mathbf{p}^2}{E} \right\rangle C_2&\equiv&  g_\chi \int \frac{d^3p}{(2\pi)^3E} \frac{\mathbf{p}^2}{E}\, C[f_\chi]\,.
\eea

Plugging in $C=C_{\rm ann}+C_{\rm el}$ as provided in Eqs.~(\ref{Cann_simp},\ref{Cresult}), finally, 
we arrive at a coupled set of equations that constitutes one of our main results:\footnote{%
\label{Aarsencomp}%
This extends the results presented in \cite{vandenAarssen:2012ag}. Compared to that reference, we have kept 
terms proportional to $Y_{\rm eq}$ (see also \cite{Duch:2017nbe}) and adopted a fully relativistic temperature 
definition in
Eqs.~(\ref{ydef},\ref{tdef}). The latter indeed turns out to be important outside the highly non-relativistic regime and
is the origin of the last term in Eq.~(\ref{yfinalfinal}), as well as  
the corrected form of $\left\langle \sigma v\right\rangle_{2}$ -- which now (unlike in its original form) can be seen as 
a proper thermal average in the sense that a constant $\sigma v_{\rm lab}$ leads to
$\left\langle \sigma v\right\rangle_{2}=\sigma v_{\rm lab}$ for {\it all} values of $T$ (i.e.~not only 
for $T\to0$).\\
We note that both $\left\langle p^4/E^3\right\rangle$ and the integral over $\epsilon_+$ can be 
expressed in terms of a series of Bessel functions when expanding $E$ in the denominator 
around $E=m$. Since this series does not converge very fast for the relatively small values of $x$
that we will be interested in here, however, we do not display these series.
} 
\bea
\frac{Y'}{Y} &=& \frac{s Y}{x \tilde H}\left[
\frac{Y_{\rm eq}^2}{Y^2} \left\langle \sigma v\right\rangle- \left\langle \sigma v\right\rangle_{\rm neq}
\right]\,, \label{Yfinalfinal}\\
\frac{y'}{y} &=&   \frac{\gamma(T)}{x\tilde H}\left[\frac{y_{{\rm eq}}}{y} -1\right]
+\frac{sY}{x\tilde H}\left[
\left\langle \sigma v\right\rangle_{\rm neq}-\left\langle \sigma v\right\rangle_{2,{\rm neq}}
\right] \label{yfinalfinal}\\ 
&&+\frac{sY}{x\tilde H}\frac{Y_{\rm eq}^2}{Y^2}\left[
\frac{y_{{\rm eq}}}{y}\left\langle \sigma v\right\rangle_{2}-\left\langle \sigma v\right\rangle
\right]
+\frac{H}{x\tilde H} \frac{\langle p^4/E^3 \rangle_{\rm neq}}{3T_\chi}\,. \nonumber
\eea
Here, in addition to $\langle \sigma v\rangle$ in Eq.~(\ref{therm_av1}), we also introduced 
another, temperature-weighted thermal average:
\bea
\label{therm_av2}
 \left\langle \sigma v\right\rangle_{2}
 &\equiv& 
 \frac{g_\chi^2}{T n_{\chi,{\rm eq}}^2} \int 
 \frac{d^3p\, d^3\tilde p}{(2\pi)^6}
 \frac{p^{2}}{3E} 
 \sigma v_{\bar\chi\chi\rightarrow \bar f f} f_{\chi,{\rm eq}}({\mathbf{p}}) f_{\chi,{\rm eq}}(\tilde{\mathbf{p}})\nonumber\\
 \\
 &=& 
 \int_{1}^\infty d\tilde s\,  \sigma_{\bar\chi\chi\rightarrow \bar f f}v_{\rm lab}
 \frac{4{\tilde s} (2\tilde s-1){x}^3}{3{K_2}^2(x)}  \nonumber\\
 &&\int_{1}^\infty d\epsilon_+ e^{-2\sqrt{\tilde s} x \epsilon_+}\,\Bigg[
 \epsilon_+\sqrt{(\tilde s-1)(\epsilon_+^2-1)}\nonumber\\
 && +\frac1{2\sqrt{\tilde s}}\log\left(
  \frac{\sqrt{\tilde s}\epsilon_+-\sqrt{(\tilde s-1)(\epsilon_+^2-1)}}{\sqrt{\tilde s}\epsilon_++\sqrt{(\tilde s-1)(\epsilon_+^2-1)}}
  \right)
 \Bigg]\,,\label{therm_av2eq}
\eea
where we have used $\epsilon_+ \equiv (E+ \tilde{E})/\sqrt{s}$.
The `out-of equilibrium average' $\left\langle \sigma v\right\rangle_{2,{\rm neq}}$ is defined as 
in Eq.~(\ref{therm_av2}), but for arbitrary  $n_{\chi}$, $f_\chi(\mathbf{p})$ -- and hence also
$1/T\to1/T_\chi$ in the normalization; the 
last equality, Eq.~(\ref{therm_av2eq}), thus does not hold in this case. Correspondingly, 
$\left\langle \sigma v\right\rangle_{\rm neq}$ is defined in analogy to Eq.~(\ref{therm_av1}),
but equals in general not the expression given in Eq.~(\ref{therm_av1eq}).

Two comments about this central result are in order. 
The first comment, more important from a practical point of view, is that the set of equations 
(\ref{Yfinal}, \ref{yfinal}) includes higher moments of $f_\chi$,
and hence does not close w.r.t.~the variables $Y$ and $y$. Concretely, we need additional input
to determine the quantities $\left\langle \sigma v\right\rangle_{{\rm neq}}$, 
$\left\langle \sigma v\right\rangle_{2,{\rm neq}}$ and $\langle p^4/E^3 \rangle_{\rm neq}$ in 
Eqs.~(\ref{Yfinalfinal}, \ref{yfinalfinal})
in terms of only $y$ and $Y$. 
We will make the following ansatz for these quantities:
\bea
  \left\langle \sigma v\right\rangle_{{\rm neq}} &=&  \left.\left\langle \sigma v\right\rangle\right|_{T=y s^{2/3}/m_\chi}, 
  \label{svansatz}\\
\left\langle \sigma v\right\rangle_{2,{\rm neq}} &=& \left.\left\langle \sigma v\right\rangle_{2}\right|_{T=y s^{2/3}/m_\chi},
\label{sv2ansatz}\\
\langle p^4/E^3 \rangle_{\rm neq} &=&
\left[\frac{g_\chi}{2\pi^2 n_{\chi,\text{eq}}(T)}\int dp\frac{p^6}{E^3} e^{-\frac{E}{T}}\right]_{T=y s^{2/3}/m_\chi}
\label{p4E3ansatz}\,.
\eea
These expressions would, in particular, result from a DM phase-space distribution of the form
\be
\label{MB_ansatz}
f_\chi = \frac{n_\chi(T)}{n_{\chi, \text{eq}}(T_{\chi})}\exp \left(-\frac{E}{T_\chi}\right) \bigg|_{T_\chi=y s^{2/3}/m_\chi}\,,
\ee
which describes a situation in which the DM particles follow a Maxwellian velocity distribution with 
a temperature different from that of the heat bath (as expected, e.g., if the DM particles exhibit significant 
self-scattering \cite{Feng:2009hw,Buckley:2009in,Feng:2010zp,vandenAarssen:2012ag}). We emphasize, 
however, that from the point of view of solving the coupled set 
of equations (\ref{Yfinalfinal}, \ref{yfinalfinal}), there is no need to make such a relatively strong
assumption about $f_\chi(\mathbf{p})$: {\it any} form of $f_\chi$ that leads to (very) similar results for the 
quantities given in Eqs.~(\ref{svansatz}) -- (\ref{p4E3ansatz}) will also lead to (very) similar results for
$Y(x)$ and $y(x)$. In other words, we expect our coupled system of Boltzmann equations to 
agree with the full numerical solution discussed in the next section
-- concerning the evolution of $Y$ and $y$ -- if and only if the ansatz in Eqs.~(\ref{svansatz},\ref{sv2ansatz},\ref{p4E3ansatz}) coincides with the 
corresponding averages numerically determined from the `true' phase-space distribution. 
As we will see later,  this is indeed very often the case.

The second comment concerns the first term
on the r.h.s.~of Eq.~(\ref{yfinalfinal}), which is proportional to the second moment of the 
elastic scattering term given in Eq.~(\ref{Cresult}). As that latter expression is valid only to lowest 
order in $p^2/E^2\sim p^2/m_\chi^2\sim 1/x$, we had for consistency also to neglect any higher-order 
corrections in these quantities to the elastic scattering part of $C_2$ when deriving our final result.
As discussed in Appendix  \ref{app:semirel}, in fact, there is no simple way of determining the 
next-to-leading order corrections to $C_{\rm el}$. If we use our default semi-relativistic 
scattering term given in Eq.~(\ref{Csemirel}), however, including the resulting corrections from sub-leading
orders corresponds to replacing in Eq.~(\ref{yfinalfinal})
\begin{align}
\label{gamma_semirel}
&T_\chi \left[\frac{y_{{\rm eq}}}{y} -1\right]=T-T_\chi\\
&\to 
T-T_\chi+\frac16\left\langle \frac{p^4}{E^3}\right\rangle
-\frac56T\left\langle \frac{p^2}{E^2}\right\rangle+\frac13T\left\langle \frac{p^4}{E^4}\right\rangle\,.\nonumber
\end{align}
By construction, see Appendix  \ref{app:semirel}, this operator must still be an attractor to the
equilibrium solution, and hence be proportional to (some power of) $T-T_\chi$; for the ansatz of 
Eq.~(\ref{MB_ansatz}), e.g., this can easily be verified directly.
In practice, this replacement has very little impact on the evolution of $Y$ and $y$, even at times as 
early as $x\sim10$. We can think of the resulting small differences as a measure of the intrinsic uncertainty 
associated to our treatment of the scattering term.

\subsection{The full phase-space density evolution}
\label{sec:magicmichael}

We now turn to solve the Boltzmann Eq.~(\ref{diff_boltzmann}) at the full phase-space density level. This is numerically 
more challenging, but allows to assess the validity of the assumptions in previous sections and to track deviations (as we 
will see can occur) from the standard Maxwell Boltzmann velocity distribution.  To achieve this, we  start by 
re-expressing Eq.~\eqref{diff_boltzmann} in the two dimensionless coordinates
$$
x(t,p)\equiv m_\chi/T   \quad \text{and} \quad q(t,p) \equiv p/T,
$$
where the monotonic temperature $T(t)$ replaces as before the time parameter $t$ via our $x(T)$, and $q$ is now the 
`momentum' coordinate that depends on both $t$ and $p$.
In these variables,  we can rewrite the Liouville operator on the  l.h.s.~of Eq.~\eqref{diff_boltzmann} as
\begin{align}
\left(\partial_t-H\mathbf{p}\cdot\nabla_\mathbf{p}\right)=\partial_t - H p \partial_p
= \tilde H \left( x\,\p_ x   - \tilde g\, q\,\p_q \right)\,.
	\label{eq:L}
\end{align}
Here, we used the fact that the system is isotropic and assumed, as in the previous sections, that entropy is conserved. 
With the collision terms given in Eqs.~(\ref{Cann_simp}) and (\ref{Csemirel}), 
the Boltzmann equation for $f_\chi$ now becomes
\bea
\p_x f_\chi(x,q)
 &=& \frac{m_\chi^3}{\tilde{H} x^4}\frac{g_{\bar \chi} }{2\pi^2}\int{\!d\tilde q\;\tilde q^2}\;  \frac{1}{2}\!\int{\!d\!\cos{\theta}\,}  \; v_{\rm M\o l}  \sigma_{\bar\chi\chi\rightarrow \bar f f} 
\nn\\
&& \times \left[ f_{\chi,{\rm eq}}(q)f_{\chi,{\rm eq}}(\tilde q)-f_\chi(q)f_\chi(\tilde q) \right] \nn\\
&+& \frac{\gamma(x)}{2 \tilde{H} x} \left[x_q \partial^2_q + \left(q+\frac{2 x_q}{ q} + \frac{q}{x_q}\right)\partial_q+3\right]f_\chi\nn\\
&+&  \tilde{g}   \frac{q}{x}  \partial_q f_\chi,
	\label{eq:BEps}
\eea
where $x_q \equiv \sqrt{x^2 +q^2}$ and $\theta$ is the angle between $\bf{q}$ and $\bf{\tilde q}$.\\[-2ex]
 
The benefits of this rewriting is two-fold.  First, the interpretation of the Boltzmann equation becomes very transparent, 
in the sense that this ``comoving'' phase space density $f_\chi(x,q)$ clearly stays unaltered for  $\tilde g(x) = 0$ and 
vanishing annihilation and scattering rates (being proportional to  $\sigma_{\bar\chi\chi\rightarrow \bar f f} $ and 
$\gamma$, respectively). The new coordinates thus absorb how momentum and DM density 
change exclusively due to the Hubble expansion. (For non-vanishing $\tilde g$, these quantities continue 
to scale in the same way with the scale factor $a$, but taking into account that $a \propto {g_{\rm eff}^s}^{-1/3}T^{-1}$).  
Second, the use of a comoving momentum $q \equiv p/T$ significantly helps numerical calculations that extend over a 
large range in $x=m_\chi/T$.  In fact, $f_\chi(x,q)$ is expected to stay unchanged in shape both in the early 
semi-relativistic and kinetically coupled regime, where $f_\chi \sim e^{-p/T_\chi} = e^{-q}$ given that $T_\chi = T$,  
as well as in the 
late non-relativistic kinetically decoupled regime, where $f_\chi \sim e^{-p^2/(2m T_\chi)}  \propto e^{-q^2/(2m)}$ 
given that $T_\chi \propto T^2$ in this case
--- at least as long as $\tilde g =0$ and the DM phase-space distribution remains close to 
Maxwellian as in Eq.~(\ref{MB_ansatz}). 

Let us stress that here, unlike for our discussion in the previous subsection, it is indeed mandatory to use the 
semi-relativistic form of Eq.~\eqref{Csemirel} for the scattering 
operator when discussing the evolution of the phase-space density, in the sense that it must drive the distribution 
function $f_\chi(q)$ towards the fully relativistic form  $\propto e^{-E/T}$ (and not as Eq.~\eqref{Cresult} to the 
non-relativistic approximation 
$\propto e^{-\frac{q^2}{2m}/{T}}$). 
The importance of this can 
be seen by comparing the second and the third line of Eq.~\eqref{eq:BEps}. The term in the second line will 
always drive DM  annihilation to occur unless an equilibrium distribution $f_{\rm eq}$ is reached.  The term in the 
third line determines towards which 
equilibrium shape the scattering operator will drive the DM distribution $f_\chi(q)$. If the scattering attractor 
distribution would not match the $f_{\rm eq}(q)$ of the second line, then scattering could artificially drive annihilation to 
occur. For more discussions of the semi-relativistic aspects of the scattering term, see Appendix \ref{app:semirel}.

We then use a technique that  discretizes  the unbounded momentum variable $q$ into a finite number of 
$q_i$ with $i\in \{1,2,\ldots,N\}$. This allows to rewrite our  \emph{integro partial differential equation}  into 
a set of $N$ coupled ordinary differential equations (ODEs):
\bea
\frac{d}{dx} f_i &=& \nn\\
&& \hspace{-0.8cm} \frac{m_\chi^3}{\tilde{H} x^4}\frac{g_{\bar \chi} }{2\pi^2}
\sum_{j=1}^{N-1} 
\frac{\Delta \tilde q_j}{2} 
\Big[
   {\tilde q_j^2 } \,\langle v_{\rm M\o l} \sigma_{\bar\chi\chi\rightarrow \bar f f} \rangle_{i, j}^{\theta}
   \left( f^{\rm eq}_i f^{\rm eq}_j \!-\! f_if_j \right) \nn\\
&+&     {\tilde q_{j+1}^2 } \,\langle v_{\rm M\o l} \sigma_{\bar\chi\chi\rightarrow \bar f f} \rangle_{i, {j+1}}^{\theta}
 \left( f^{\rm eq}_i f^{\rm eq}_{j+1} \!-\! f_if_{j+1} \right)
\Big]
  \nn\\
&+&  \frac{\gamma(x)}{2 \tilde{H} x} \left[x_{q,i} \partial^2_q f_i+ \left(q_i\!+\!\frac{2 x_{q,i}}{q_i} \!+\! \frac{q_i}{x_{q,i}}  \right)\partial_q f_i+3 f_i\right] \nn\\
&+&    \tilde{g}   \frac{q_i}{x}  \partial_q f_i,
\label{eq:BE_num}
\eea
where $f_i \equiv f_\chi(x,q_i)$, and the derivatives $\p_q f_i$ and $\p^2_q f_i$ are determined numerically 
by finite differentials using several neighboring points to $f_i$.
$\langle v_{\rm M\o l} \sigma_{\bar\chi\chi\rightarrow \bar f f} \rangle_{i, j}^{\theta}$ is the
velocity-weighted cross section averaged over $\theta$ (which is evaluated analytically or numerically) 
as a function of $q_i$ and $\tilde{q}_j$, and $\Delta \tilde q_j \equiv  \tilde{q}_{j+1}- \tilde{q}_j$.  Finally, 
the DM number density in Eq.~\eqref{eq:n} is determined by trapezoidal integration.

Numerous numerical tests have been performed to ensure stability of our solutions to the ODEs of 
Eq.~\eqref{eq:BE_num} and that 
imposed conditions on the now emerged boundary points  (at $q_1$ and $q_N$) are physically sound.  
It turns out that very small stepsizes over a large range in $q$ are required for solving these stiff ODEs. 
We typically used the 
range $q_1 =10^{-6}$ to $q_N=50$ with about thousand steps in between, and set the two last terms of 
Eq.~\eqref{eq:BE_num} to zero at $q_N$ while using forward derivatives to evaluate them at $q_1$. 
By the use of  the ODE15s code in {\sf MatLab},
 and by analytically deriving internally required  Jacobians, we are 
able to efficiently calculate 
the full phase-space evolution for the freeze-out after optimizing numerical settings. On the time scale of a few minutes 
we can derive the relic abundance for a given DM model.
 The code is general enough to be adapted to any standard single WIMP setup.

\section{Scalar Singlet Dark Matter}
\label{sec:singlet}

The simplest example of a renormalizable model providing a WIMP DM candidate is 
the Scalar Singlet model \cite{Silveira:1985rk,McDonald:1993ex,Burgess:2000yq},
originally proposed as DM made of `scalar phantoms' by Silveira and Zee \cite{Silveira:1985rk}.
In this model, the only addition to the standard model is a real gauge-singlet scalar field $S$
which is stabilized by a $\mathbb{Z}_2$ symmetry and never obtains a non-vanishing vacuum 
expectation value. The simplicity of the model has in itself triggered considerable interest  
\cite{Yaguna:2008hd,Profumo:2010kp,Arina:2010rb,Mambrini:2011ik,Lerner:2009xg,Herranen:2015ima,Kahlhoefer:2015jma,Profumo:2007wc,Barger:2008jx,Cline:2012hg}, with a further boost of attention after the discovery of the Higgs boson \cite{Djouadi:2011aa,Cheung:2012xb,Endo:2014cca,Djouadi:2012zc,Cline:2013gha,Urbano:2014hda,He:2016mls,Escudero:2016gzx,Goudelis:2009zz,Craig:2014lda,Han:2016gyy,Ko:2016xwd,Beniwal:2015sdl,Cuoco:2016jqt}. 
Recently, the GAMBIT \cite{Athron:2017ard} collaboration presented the so far most comprehensive 
study of this model by performing a global fit taking into account experimental constraints from 
both direct, indirect and accelerator searches for DM \cite{Athron:2017kgt}. 

Interestingly, the resulting parameter region with the highest profile likelihood in this global fit is the one where
the Scalar Singlet mass $m_S$ is about half that of the SM Higgs mass, $m_h$, and where the DM 
abundance today is set by the resonant annihilation of two DM particles through an almost on-shell 
Higgs boson. As we will see, it is exactly in this parameter region that the standard way of 
calculating the relic density, as implemented in all previous studies of this model, fails because
kinetic decoupling happens so early that it essentially coincides with chemical decoupling.
Instead, the formalism introduced in the previous section provides a reliable calculation of
the relic abundance of Scalar Singlet DM.

\subsection{Model setup}

The model symmetries, along with the requirement of renormalizability, uniquely determine the 
form of the Lagrangian to be 
\begin{equation}
{\cal L}_{\rm SZ} = {\cal L}_{\rm SM} + \frac{1}{2} \partial_\mu S \partial^\mu S - \frac{1}{2} \mu_S^2 S^2 - \frac{1}{2} \lambda_S S^2 H^\dagger H - \frac{1}{4!} \lambda_{SS} S^4,
\end{equation}
where $H$ is the SM Higgs doublet. The $S$ boson mass receives contributions from both the bare mass  
term, $\mu_S$,  and from electroweak symmetry breaking, leading to 
$m_S = \sqrt{\mu_S^2 + \frac{1}{2}{\lambda_S v_0^2} }$, 
where $v_0=246.2$~GeV is the Higgs vacuum expectation value. We adopt the Higgs mass and the total width 
from decay to SM particles to be 
$m_h=125.09$~GeV \cite{Patrignani:2016xqp} and $\Gamma_{h, SM} = 4.042$~MeV \cite{Dittmaier:2011ti}. For the moment, we neglect the quartic 
self-coupling $\lambda_{SS}$, but will later comment on its potential (minor) impact on relic density calculations. 

The annihilation cross section of DM pairs to SM particles, apart from $hh$ final states, is given by \cite{Cline:2013gha}
\begin{equation}
\sigma v_{\rm CMS} = \frac{2 \lambda_S^2 v_0^2}{\sqrt{s}} \, |D_h(s)|^2 \, \Gamma_{h\to \rm{SM}}(\sqrt{s})\,,
\label{eq:sigmav}
\end{equation}
where $\Gamma_{h\to \rm{SM}}(\sqrt{s})$ is the partial decay width of a Standard-Model Higgs boson of 
mass $\sqrt{s}$, and 
\begin{equation}
\label{hdenom}
|D_h(s)|^2 = \frac{1}{(s-m_h^2)^2+m_h^2 \Gamma_h^2 }.
\end{equation}
The total Higgs width $\Gamma_h$ in the above propagator, but not elsewhere, includes not only all 
SM channels but also the $h\to SS$ channel if it is open. For 
$\Gamma_{h\to \rm{SM}}(\sqrt{s})$, as in \cite{Cline:2013gha}, we use tabulated 
values for $\sqrt{s} < 300$~GeV from \cite{Dittmaier:2011ti} and analytic expressions at higher $\sqrt{s}$. 
Note however that the latter high $\sqrt{s}$ 
region has no impact on the relic density in the studied Scalar Singlet mass range. Likewise, 
the channel $SS\to hh$ lies outside our kinematic region of interest.

For the elastic scattering processes, we take into account DM scattering with all SM fermions. Being mediated 
only by a Higgs in the $t$-channel, the corresponding squared amplitude takes a particularly simple form,
\be
\left| \mathcal{M}_{Sf\to Sf}\right|^2 = 2{N_f\lambda_S^2m_f^2}\frac{4m_f^2-t}{(t-m_h^2)^2} \,,
\label{eq:Mscatt_t0}
\ee
where $m_f$ is the mass of the SM fermion and the color factor is $N_f= 3$ for quarks and $N_f= 1$ 
for leptons.  Averaging over the transferred momentum, as in Eq.~(\ref{eq:sigmaT}), 
we thus find
\bea
\left\langle \left| \mathcal{M}\right|^2 \right\rangle_t=&&
\sum_f\frac{N_f\lambda_S^2m_f^2}{2k^4}\Big[
\frac{2k^2_\mathrm{cm}-2m_f^2+m_h^2}{1+m_h^2/(4k^2_\mathrm{cm})} \nonumber \\
&&-\left( m_h^2-2m_f^2\right) \log\left(1+4k^2_\mathrm{cm}/m_h^2 \right)
\Big]\,.
\label{eq:Mscatt}
\eea
Note that the sum here runs over all relevant fermions and antifermions {\it separately}.

The hierarchical Yukawa structure of the Higgs couplings leads to the scattering rate being 
dominated by the interactions with the heaviest fermions that for a given temperature are still 
sufficiently abundant in the plasma. In the range of DM masses $m_S$ that 
we are interested in, freeze-out happens around $T\sim \mathcal{O}(1\, \rm{GeV})$, which 
is not far from the temperature of the QCD phase transition. 
Consequently, the details of this transition and the SM plasma can have a significant impact 
on the scattering rate; a study which goes beyond the scope of this work. Therefore, we follow the
literature and adopt two extreme scenarios that can be thought of as bracketing the actual size of 
the scattering term:
\begin{description}
\item[A] all quarks are free and present in the plasma down to temperatures of $T_c = 154$\,MeV (largest scattering 
scenario, as adopted in \cite{Gondolo:2012vh})
\item[B] only light quarks ($u$, $d$, $s$) contribute to the scattering, and only for temperatures above 
$4T_c \sim 600$\,MeV, below which hadronization effects start to become sizeable \cite{Boyanovsky:2006bf} (smallest scattering scenario, as adopted in \cite{Bringmann:2009vf}).
\end{description}

Finally, we adopt the recent results from Drees et al.~\cite{Drees:2015exa} for the effective number of relativistic
degrees of freedom $g_{\rm eff}(T)$ that enter the calculation of the Hubble rate during radiation domination,
 $H=\sqrt{4\pi^3g_{\rm eff}/45}\,T^2/m_{\rm Pl}$, as well as the entropy degrees of freedom entering for example in the 
 calculation of $\tilde g(T)$ as defined in Eq.~(\ref{gtildedef}).

\subsection{Relic density of scalar singlet dark matter}
\label{sec:singlet_rd}

\begin{figure}[t!]
  \includegraphics[width=0.9\columnwidth]{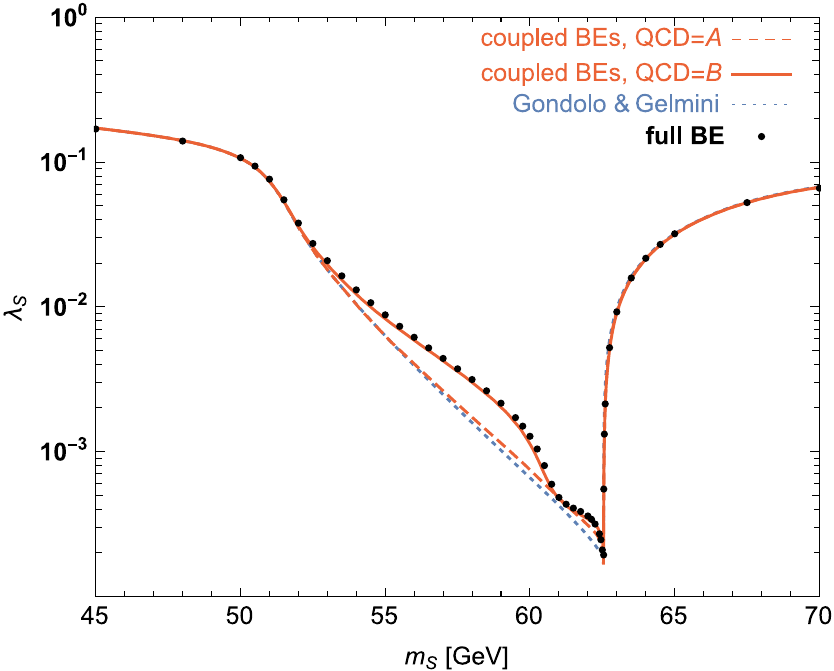}
  \caption{The required value of the Singlet-Higgs coupling $\lambda_S$, as a function of the Scalar 
  Singlet mass $m_S$, in order to obtain a relic density of  $\Omega h^2=0.1188$.  The blue dashed
  line shows the standard result as established by Gondolo \& Gelmini \cite{Gondolo:1990dk}, based 
  on the assumption of local thermal equilibrium during freeze-out. 
  For comparison, we also plot the result of solving instead the coupled system of Boltzmann equations 
  \eqref{Yfinalfinal} and \eqref{yfinalfinal} for the maximal (`B') and minimal (`A') quark scattering scenarios 
  defined in the main text (red solid and dashed lines, respectively). Finally, we show the result of fully solving
 the Boltzmann equation numerically, for the maximal quark scattering scenario and with no DM self-interactions 
 included (`full BE').
  }
    \label{fig:RDcont1}
\end{figure}

Let us first compute the relic density following the standard treatment adopted in the literature. 
To this end, we numerically solve Eq.~(\ref{boltzYsimp}) for a given set of parameters $(m_S,\lambda_S)$ 
and determine the resulting asymptotic value of $Y_0$. The blue dashed line in Fig.~\ref{fig:RDcont1} 
shows the contour in this plane that results in $Y_0$ corresponding to a relic density of 
$\Omega h^2=0.1188$, c.f.~Eq.~(\ref{eq:oh2}).
We restrict our discussion to values of $m_S$ in the kinematic range where $\langle \sigma v\rangle$ is 
enhanced due to the Higgs propagator given in Eq.~(\ref{hdenom}), and
the coupling $\lambda_S$ that results in the correct relic density is hence correspondingly decreased.
This curve agrees with the corresponding result obtained in Ref.~\cite{Cline:2013gha}. 

For comparison, we show in the same figure the required value of $\lambda_S$ that results when instead
solving the coupled system of Boltzmann equations \eqref{Yfinalfinal} and \eqref{yfinalfinal},
or when numerically solving the full Boltzmann equation as described in Section \ref{sec:magicmichael} .
Here, the solid (dashed) line shows the situation for the `B' (`A') scenario for scatterings on quarks.
Outside the resonance region, the coupled Boltzmann equations lead to identical results compared to 
the standard approach, indicating
that kinetic decoupling indeed happens much later than chemical  decoupling and that the assumption 
of local thermal equilibrium during chemical freeze-out thus is satisfied.
For DM masses inside the resonance region,  on the other hand, we can see that the two methods can 
give significantly different results, implying that this assumption must be violated. For the same reason,
a smaller scattering rate (as in scenario `B') leads to an even larger deviation from the 
standard scenario than the maximal scattering rate adopted in scenario `A'.

\begin{figure}[t!]
\vspace{0.11cm}
  \includegraphics[width=0.91\columnwidth]{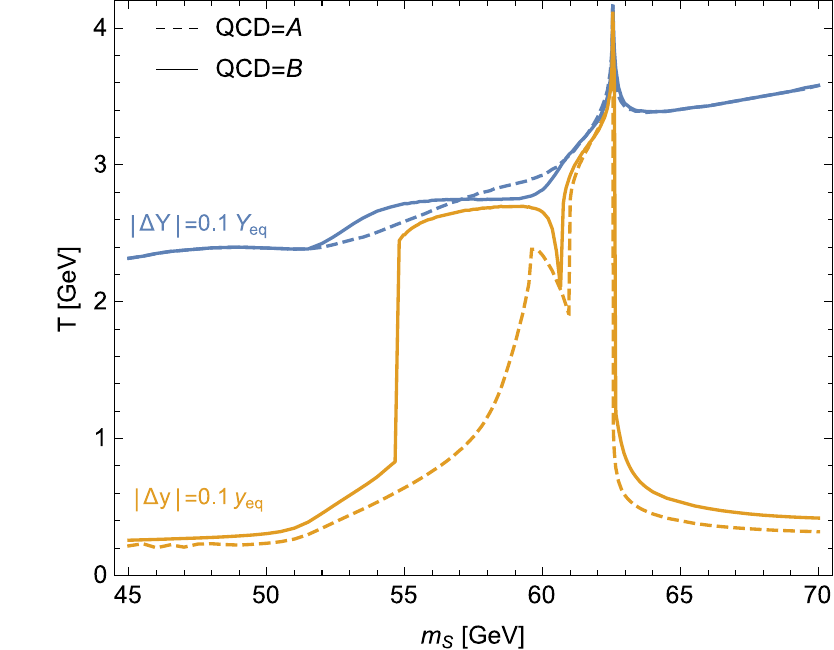}
  \caption{Temperatures at which DM number density and velocity dispersion (`temperature') start to deviate from their
  equlibrium values, defined for the purpose of this figure as $|Y-Y_{\rm eq}|=0.1\,Y_{\rm eq}$ and $|y-y_{\rm eq}|=0.1\,y_{\rm eq}$, respectively.
  These curves are based on solving  the coupled system of Boltzmann equations  \eqref{Yfinalfinal} and 
  \eqref{yfinalfinal},  for the same parameter combinations as in Fig.~\ref{fig:RDcont1} (resulting thus in the correct 
  relic density). }
    \label{fig:TkdTcd}
\end{figure}

This interpretation is explicitly confirmed in Fig.~\ref{fig:TkdTcd}, where we plot the temperatures at 
which the DM number density and temperature start to deviate from the equilibrium values:
in the parameter range that we focus on here, kinetic decoupling happens indeed very close to 
chemical decoupling. The reason for this very early kinetic decoupling  is straight-forward 
to understand as the result of a strongly suppressed momentum transfer rate $\gamma(T)$, 
compared to the annihilation rate, due to two independent 
effects: {\it i)} the small coupling $\lambda_S$ needed to satisfy the relic density
requirement, without a corresponding resonant enhancement of $\gamma(T)$, and {\it ii)} the scattering 
rate being proportional to the Yukawa coupling squared,
which favours scattering with Boltzmann-suppressed heavy fermions. We note that the 
latter point also explains the relatively large difference between the two extreme quark scattering 
scenarios used here for illustration (in scenario `B', the largest Yukawa couplings do not contribute
to the scattering). 

\begin{figure}
  \includegraphics[width=0.9\columnwidth]{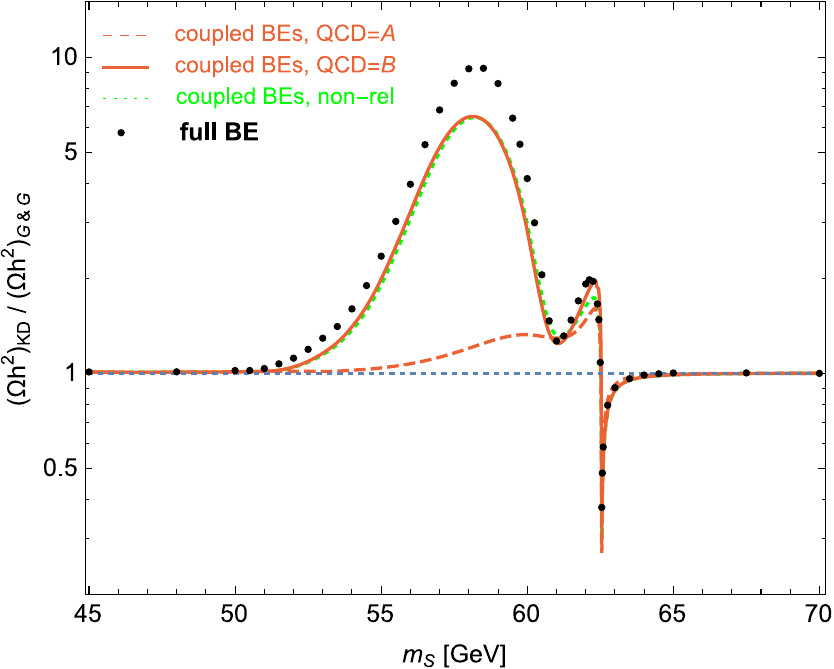}
  \caption{The impact of the improved treatment of the kinetic decoupling on the relic density for parameter
  points that would satisfy the relic density constraint in the standard approach (dotted line in Fig.~\ref{fig:RDcont1}),
  both for the minimal (solid) and maximal (dashed) scenario for scattering with quarks.
 The numerical result (`full BE') implements minimal quark scattering; note that this does not take 
 into account the effect of DM self-interactions (while the other curves are consistent with assuming a
 maximal self-scattering rate). The green dashed curve shows the impact of implementing the elastic 
 scattering term in the highly non-relativistic limit, c.f.~Eq.~(\ref{Cresult}).
  }
    \label{fig:effect}
\end{figure}

In order to emphasize  the importance of our improved treatment of the decoupling history, 
we plot in Fig.~\ref{fig:effect} also the {\it ratio} of the resulting relic density to that of the
standard approach (for parameter values satisfying the relic density constraint for the latter, 
i.e.~corresponding to the  blue dashed curve in Fig.~\ref{fig:RDcont1}). Let us stress that, 
compared to the observational uncertainty in this quantity of about 1\,\%, these corrections
are by no means small even in the minimal scattering scenario `A'. In the same figure, we 
also compare our result for the coupled 
system of Boltzmann equations \eqref{Yfinalfinal} and \eqref{yfinalfinal} to the full 
numerical solution of the Boltzmann equation in phase space, as described in
Section \ref{sec:magicmichael} (black dots). Before getting back to these results,
let us briefly comment on the green dashed line in Fig.~\ref{fig:effect}, which implements 
the highly non-relativistic scattering term $C_{\text{el}}$ of Eq.~(\ref{Cresult}), and hence {\it not} the 
replacement (\ref{gamma_semirel}) in Eq.~(\ref{yfinalfinal}) which we otherwise adopt
as our default. Clearly, the impact of this choice is very limited for this approach. 
We note that the quantitative importance of the relativistic correction term proportional to 
$\langle p^4/E^3\rangle$ in Eq.~(\ref{yfinalfinal}) lies in the same ballpark, affecting the
relic density by at most $\sim$$10\%$ in the region very close to the resonance
(and below the percent-level elsewhere).

\begin{figure*}
  \includegraphics[width=0.9\columnwidth]{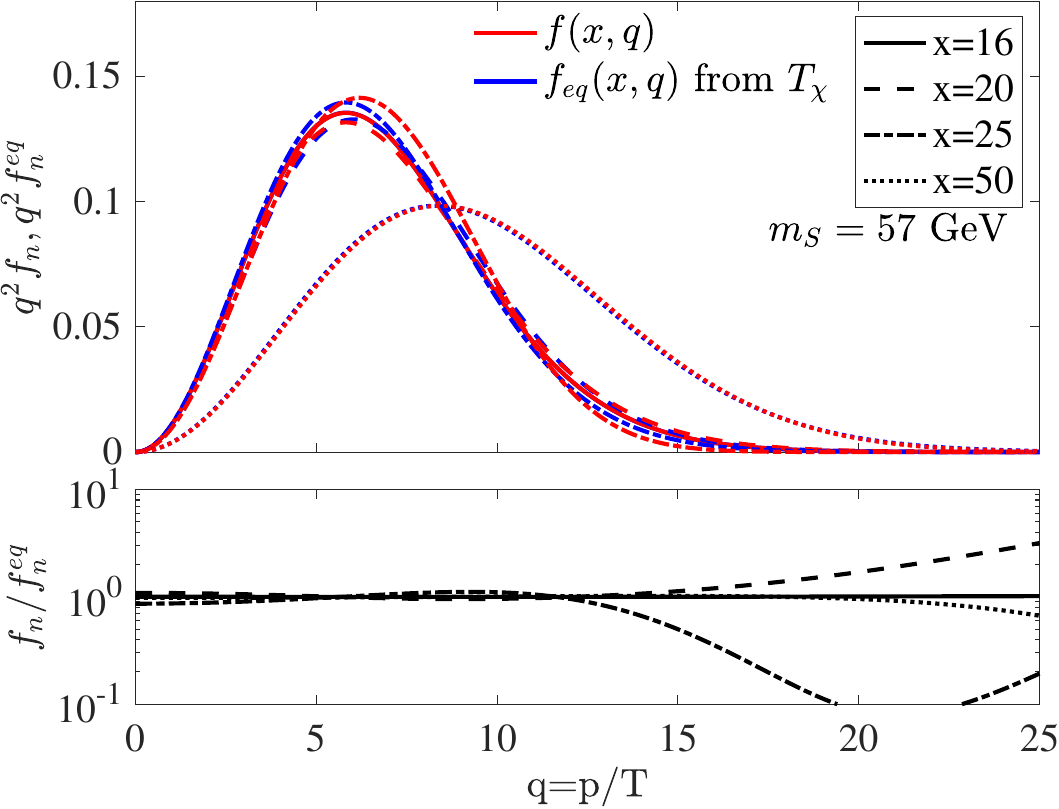}
  \hspace*{2.5cm}
  \includegraphics[trim={3.25cm 1.05cm 3.25cm 1.5cm}, width=0.53\columnwidth]{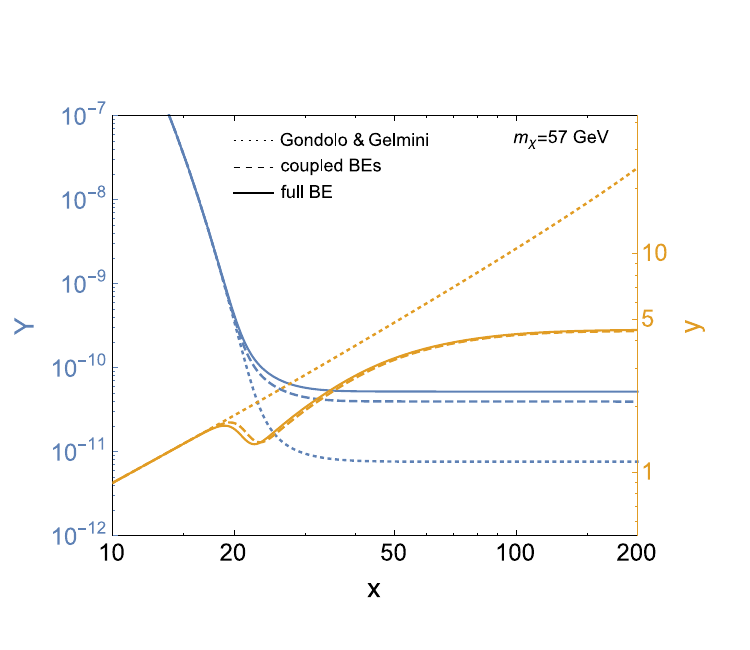}
  \hspace*{1.5cm}
  \caption{Phase space distributions and their evolution for  a Scalar Singlet DM particle with $m_S=57$\,GeV. 
  {\it Left panel:} Unit normalized phase-space distributions $f_n(q)$ from our full numerical solution of the
  Boltzmann equation 
  (red lines) and thermal equilibrium distributions  $f^\text{eq}_n(q)$ (blue lines) at four different 
  temperatures $x = m_S/T = $ 16 (solid), 20  (dashed), 25 (dot-dashed) and 50 (dotted).  The equilibrium 
  distributions  $f^\text{eq}_n$ are Maxwell-Boltzmann distributions evaluated at the 
  `temperatures' $T_\chi$, as defined in  Eqs.~(\ref{ydef}, \ref{tdef}). The bottom part shows 
  the fractional deviation from the respective thermal distribution $f_n(q) / f_n^\text{eq}(q)$. 
  {\it Right panel:} The evolution of $Y$ (blue) and $y$ (yellow), assuming a Higgs-scalar coupling that 
  leads to the correct relic density in the standard approach (dotted line in Fig.~\ref{fig:RDcont1}).
  We show these curves for the standard case (dotted lines), the approach using coupled Boltzmann 
  equations (dashed) and the full numerical result (solid). The thin gray line indicates the asymptotic 
  value of $Y$ corresponding to the observed relic density.
}
    \label{fig:PSdist}
\end{figure*}

In Appendix \ref{app:SingletDetails} we discuss in depth the time evolution of both
the coupled Boltzmann equations and the full phase-space density in the resonance
region. Let us here just mention that the characteristic features of the curves displayed in 
Figs.~\ref{fig:TkdTcd} and \ref{fig:effect}
can indeed all more or less directly be understood in terms of the highly enhanced annihilation rate in a 
relatively narrow kinematic region around the resonance, $\sqrt{s}\sim m_h\pm\Gamma_h$.
As the full numerical solution reveals, furthermore, the shape of $f_\chi(p)$ can in some cases be quite 
different from the Maxwell-Boltzmann form (\ref{MB_ansatz}) that is consistent with the
coupled system of Boltzmann equations \eqref{Yfinalfinal} and \eqref{yfinalfinal}. 
Whether this has a noticeable impact on the resulting relic density (like for $m_S\sim57$\,GeV) 
or not (like for $m_S\sim m_h/2$) again mostly depends on whether or not the shape is affected for 
those momenta that can combine to  $\sqrt{s}\sim m_h$ during chemical freeze-out. 

For illustration, we pick a DM mass of $m_S=57$\,GeV and show in Fig.~\ref{fig:PSdist} the full 
phase-space distribution for  a few selected values of $x$ (left panel) as well as the relevant 
evolution of $Y$ and $y$ (right panel). For models with DM masses in this range, 
the relatively large difference between full solution and coupled equations (as visible in 
Fig.~\ref{fig:effect}) can mostly be understood in terms of the {\it dip} in the {\it ratio} of DM 
phase-space distributions at intermediate values of $q=p/T$ that starts to develop for $x\gtrsim20$.
Concretely, the fact that the actual distribution for those momenta is slightly suppressed 
compared to a distribution fully characterized only by its second moment, as in Eq.~(\ref{MB_ansatz}),
 causes the DM particles to annihilate less efficiently, 
 $\langle \sigma v\rangle_\mathrm{neq} < \langle \sigma v\rangle$, because this is
 the momentum range probed by the resonance for these $x$ values.  
 This in turn leads  to the DM particles falling out of chemical 
equilibrium earlier, and hence a larger asymptotic value of $Y$. The reason for this momentum
suppression to develop in the first place is also to be found in the particularly efficient annihilation
close to the resonance, which leads to a depletion of DM particles with corresponding momenta
because the scattering rate is no longer sufficiently large to redistribute the phase-space distribution
to a thermal shape. We note that the bulk part of this effect is actually 
well captured 
by the coupled Boltzmann system, c.f.~
the dashed vs.~solid lines 
in the right  panel of Fig.~\ref{fig:PSdist}. For further details, 
we refer again to Appendix \ref{app:SingletDetails}.

\section{Discussion}
\label{sec:disc}

From the above discussion, we have learned that very early kinetic decoupling is not just
a theoretical possibility. It can appear in simple WIMP models, like the Scalar Singlet case,
and affect the DM relic density in a significant way. We note that the size of the latter effect is,
as expected, directly related to the size of the momentum exchange rate and hence to just
{\it how} early kinetic decoupling happens compared to chemical decoupling. Let us stress 
that,  from a  general point of view, this is 
a much more important message connected to our choice of considering two scattering scenarios
than the question of which of those scenarios is more realistic for the specific model we
have studied here.

We have also seen that the coupled system of Boltzmann equations \eqref{Yfinalfinal} and 
\eqref{yfinalfinal} provides a qualitatively very good description for the resulting DM
abundance, see in particular Fig.~\ref{fig:RDcont1}, even though for high-precision results it 
seems mandatory to actually solve
the full Boltzmann equation in phase space. As discussed in Appendix \ref{app:SingletDetails}, 
differences can arise when the true phase-space distribution is not of the Maxwellian form 
assumed in Eq.~(\ref{MB_ansatz}) -- though the two methods {\it can} actually still give almost 
identical results for the relic abundance even  when the two distribution differ vastly.
The question of under which conditions the coupled system of equations provides an accurate description
of the relic density is thus a somewhat subtle one, and requires a careful discussion of the velocity
dependence of the annihilation term in the Boltzmann equation.

An exception to this general complication is a DM {\it self}-interaction rate large enough to 
force the DM distribution into the form given by 
Eq.~(\ref{MB_ansatz}) \cite{Feng:2009hw,Buckley:2009in,Feng:2010zp,vandenAarssen:2012ag} and hence
render the coupled system of Boltzmann equations \eqref{Yfinalfinal} and 
\eqref{yfinalfinal} {\it exactly} correct (up to, as discussed, corrections due to quantum statistics).
Sizeable self-scattering rates can for example arise due to corresponding contact 
interactions, like the quartic coupling $\lambda_{SS}$ in the Scalar Singlet case, or by adding light
mediators that couple to the DM particle (which was indeed the first time such a coupled system
of Boltzmann equations was considered \cite{vandenAarssen:2012ag}, albeit in a different context).  
For the case of resonant annihilation, furthermore, the same resonance also mediates an enhanced 
self-interaction. For future work, it would hence be worthwhile to extend our numerical framework 
to even include those DM self-interaction processes. For the Scalar Singlet case, in particular, we 
expect that adding the process $SS\to h^*\to SS$ would bring all numerical 
results for the full Boltzmann equation -- e.g.~those shown in Fig.~\ref{fig:effect} -- even closer to those 
resulting from the coupled system of Boltzmann equations.

Let us finally stress that both the coupled Boltzmann equations and the numerical setup that
we have described here are very general, and can be used to consistently study early kinetic 
decoupling for a much larger range of models than the Scalar Singlet case. Obvious 
applications are other scenarios where resonant annihilation and/or annihilation to heavy 
final states is important in setting the relic abundance, see also Ref.~\cite{Duch:2017nbe}.
Further examples where the ratio of the scattering rate to the annihilation rate can be smaller than
usual, hence potentially leading to early kinetic decoupling, include Sommerfeld-enhanced 
annihilation \cite{Dent:2009bv,Zavala:2009mi,Feng:2010zp,vandenAarssen:2012ag} (if the light mediators are 
not abundant enough to take part in the 
scattering process) and annihilation to DM bound states \cite{Feng:2009mn,vonHarling:2014kha}. 
Quite in general, our methods provide a powerful means to check whether the DM particles 
are indeed in local thermal equilibrium with the heat bath around the time when their 
abundance freezes out -- which is the usual assumption, though rarely explicitly tested,  
not only in WIMP-like scenarios but also when so-called semi-annihilations \cite{DEramo:2010keq} 
are important in setting the relic density, when computing the relic abundance for 
modified expansion histories \cite{DEramo:2017gpl, Redmond:2017tja}, or in scenarios that go 
beyond simple $2\to2$ annihilation processes \cite{Hochberg:2014dra,Hochberg:2014kqa,Kuflik:2015isi}.

\section{Conclusions}
\label{sec:conc}

The standard 
way of calculating the thermal relic density of self-annihilating DM particles
rests on the assumption of local thermal equilibrium during freeze-out, and that hence
kinetic decoupling occurs much later than chemical decoupling. Here, we 
demonstrated for the first time that departure from kinetic equilibrium can instead
happen much earlier, even {\it simultaneously} with the departure from chemical 
equilibrium.

By introducing a coupled system of equations for the DM number density and its
`temperature', or rather velocity dispersion, we improved the standard way of calculating the
relic density in such cases. For an even higher accuracy in predicting the DM abundance, 
we also found a way of solving the full Boltzmann equation numerically. The latter approach
has the additional advantage of obtaining the full phase-space distribution, rather than 
only the number density, which in particular allows to test in detail the assumption of a 
Maxwellian velocity distribution adopted in the standard approach.
A numerical solver for the coupled system of Boltzmann equations, Eqs.~(\ref{Yfinalfinal},
\ref{yfinalfinal}), will be available in an upcoming version of \ds\ \cite{ds6} and
our implemented solver for the full Boltzmann equation at the phase-space 
level, Eq.~\eqref{eq:BE_num}, will be released separately.\footnote{%
Please contact any of the authors if you need these numerical routines prior to their
public release.
}

Applied to the simplest renormalizable WIMP model -- the Scalar Singlet, extensively
discussed in the literature -- we somewhat surprisingly found that the relic abundance 
predicted in the standard approach can differ by up to an order of magnitude 
from the correct treatment presented in this paper. This is rather remarkable not only in view 
of the simplicity of this model, but also because the affected region in parameter space 
happens to coincide with the best-fit region resulting from most recent global scans. 
We thus expect our results to have a noticeable phenomenological impact, and 
that our treatment will prove useful also when applied to other examples of relic density 
calculations in cases where the standard assumption of local thermal equilibrium during 
freeze-out is not exactly satisfied.

\bigskip
{\it Note added.} While preparing this work, we became aware of a dedicated study 
on resonant DM annihilation~\cite{Duch:2017nbe}, which also found that DM
can kinetically decouple much earlier than usual in this case.

\vfill
\paragraph*{Acknowledgments.---}
We thank Mateusz Duch, Joakim Edsj\"o, Bohdan Grz\c{a}dkowski, Andreas Hohenegger
and  Ayuki Kamada for very useful conversations during the preparation of this work. 
We are also grateful to Tomohiro Abe for pointing out a previously missing factor of 4 in 
Eqs.~(\ref{eq:Mscatt_t0},\ref{eq:Mscatt}), as well as our detailed discussions around 
the coupled system of Boltzmann equations (\ref{Yfinalfinal},\ref{yfinalfinal}).
AH is supported by the University of Oslo through the Strategic Dark Matter Initiative (SDI). 
MG and T.~Binder have received funding from the European Union’s Horizon 2020 research 
and innovation programme under grant agreement No 690575 and No 674896. T.~Binder 
gratefully acknowledges financial support from the German Science Foundation (DFG RTG 1493).

\appendix

\begin{figure*}[t!]
  \includegraphics[width=0.66\columnwidth]{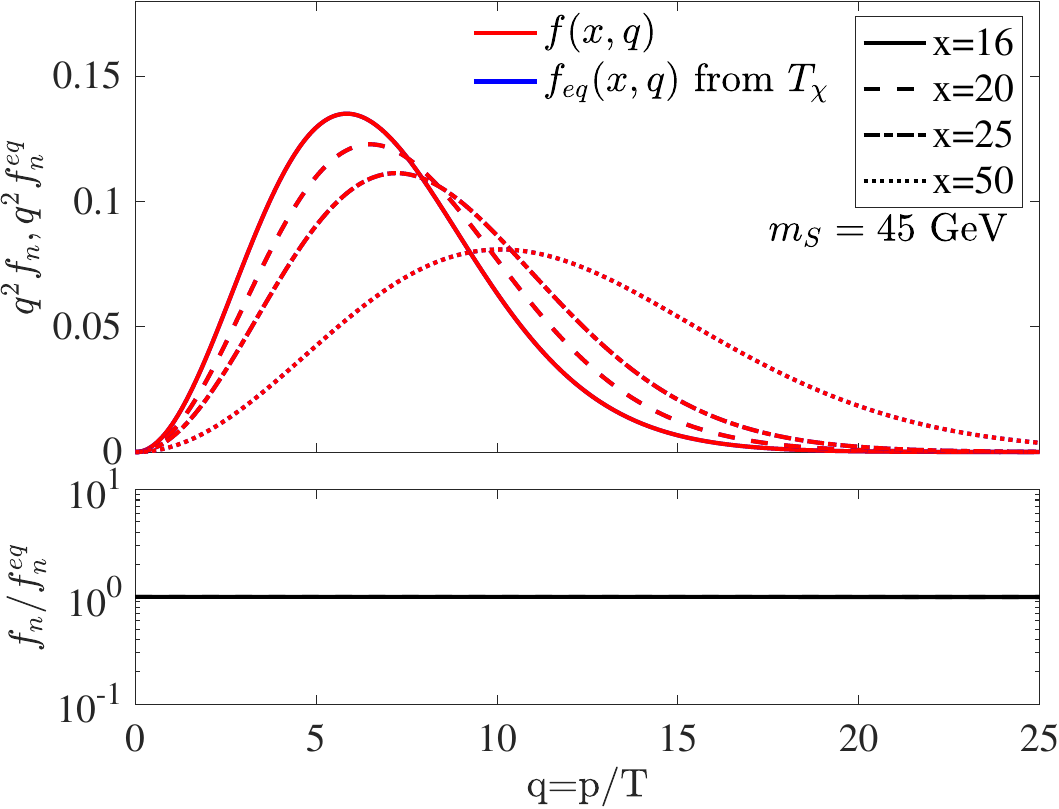}
  ~~\includegraphics[width=0.66\columnwidth]{phasespace_distr_mDM57_kCM_new}
  ~~\includegraphics[width=0.66\columnwidth]{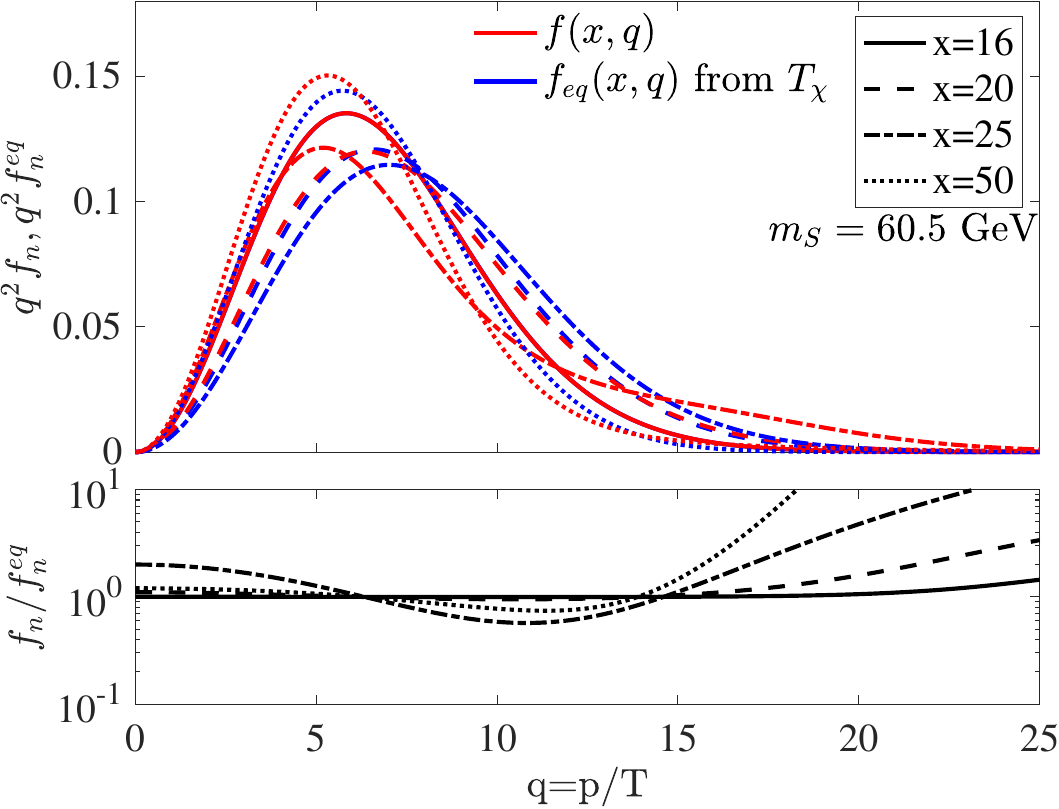}\\
  \includegraphics[width=0.66\columnwidth]{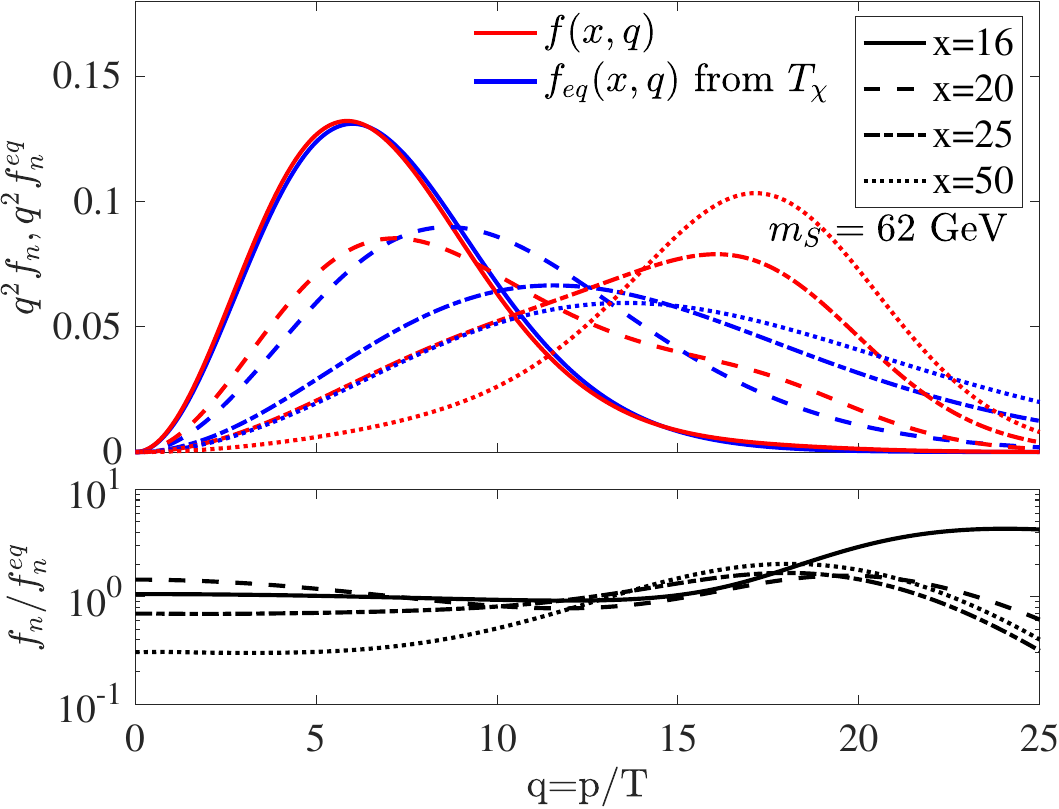}
  ~~\includegraphics[width=0.66\columnwidth]{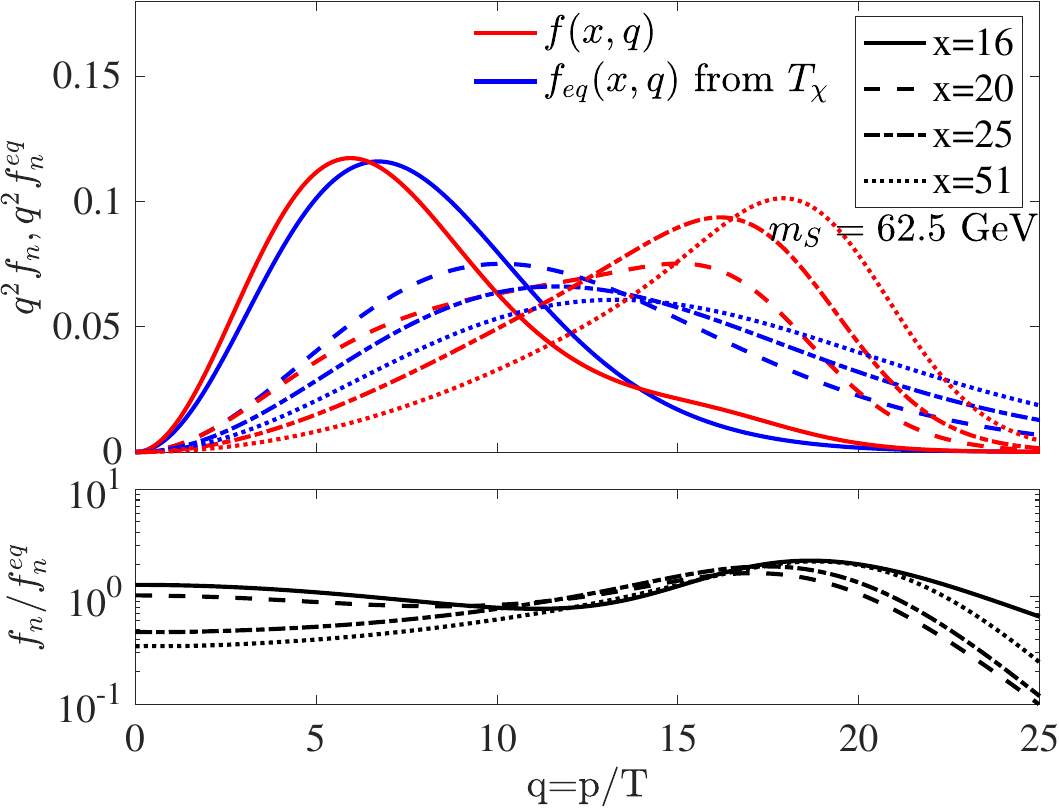}
  ~~\includegraphics[width=0.66\columnwidth]{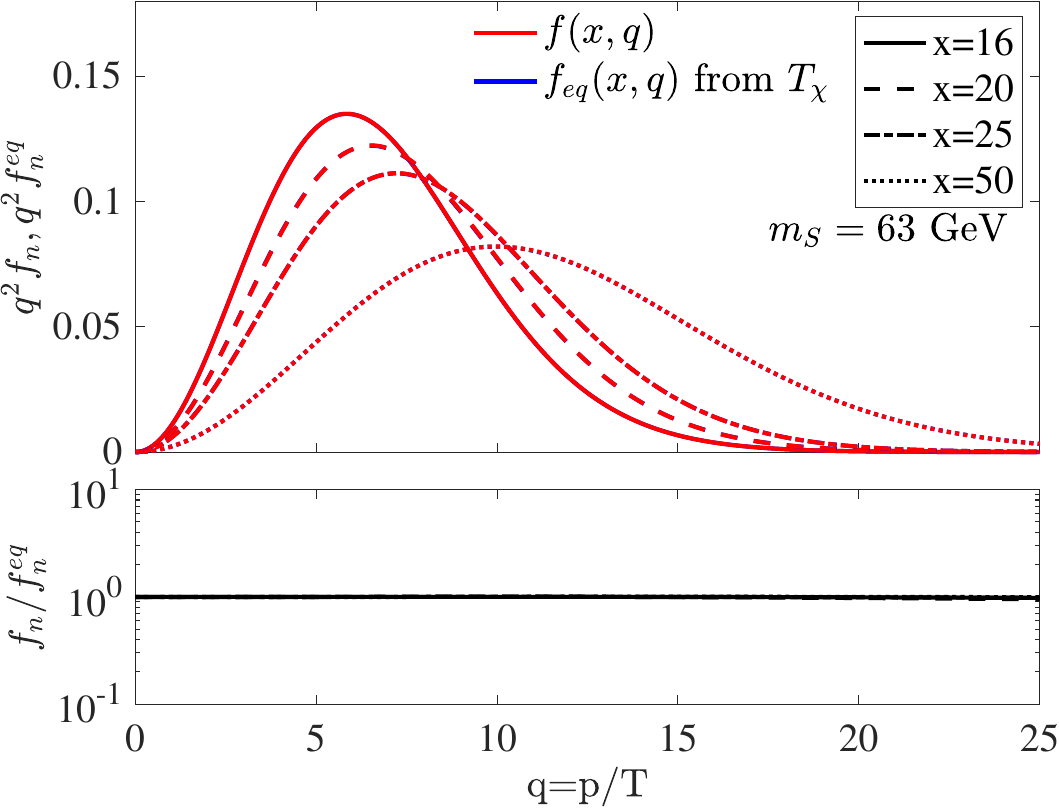}

  \caption{Same as the left panel of Fig.~\ref{fig:PSdist} in the main text, but now for comparison for various DM
  masses $m_S=45,57,60.5, 62,62.5,63$\,GeV. Note that for cases where the equilibrium 
  distributions appear to be missing in the top panels, it is just because it agrees very well 
  with the actual phase-space distribution (as also visible in the fractional deviation plotted in
  the bottom panels).  
 }
    \label{fig:PSdist_multi} \vspace{-0.4cm}
\end{figure*}
\newpage
\section{Phase-space density evolution of the Scalar Singlet}
\label{app:SingletDetails}

In Section \ref{sec:singlet_rd}, we investigated the impact of our improved treatment of the 
Boltzmann equation on the expected DM relic abundance in the Scalar Singlet model.
Here, we supplement this by discussing in some more detail the evolution of the DM
phase-space density. The main focus of this discussion, however, will be a more thorough qualitative 
understanding of the specific features seen in Fig.~\ref{fig:TkdTcd} and Fig.~\ref{fig:effect},
and the underlying interplay of chemical and early kinetic decoupling.
Specifically, we can distinguish three mass regimes:

\begin{enumerate}
\item A regime with $53\,{\rm GeV}\lesssim m_S\lesssim60.5$\,GeV, which we will refer to 
as {\it sub-resonant } because $f_\chi$ starts to deviate from its equilibrium value, 
$f_{\chi,{\rm eq}}=\exp(-E/T)$, at a temperature where the typical DM momenta are too small
to hit the resonance, i.e.~$\sqrt{s}\lesssim m_h -\Gamma_h$. As a result, we have 
$\langle \sigma v\rangle_{\rm (neq)}<\langle \sigma v\rangle_{2,{\rm (neq)}}$\footnote{%
For the sake of better readability, we will suppress the subscript `{$\rm neq$}' for 
the remainder of this section. We note that, once chemical decoupling has started, 
the contribution of thermal averages without this subscript is suppressed by a
factor of $Y_{\rm eq}/Y$ in Eqs.~(\ref{Yfinalfinal}, \ref{yfinalfinal}).
}
during the whole freeze-out process in this regime --- this is because $p^2f_\chi(p)$ peaks at 
a higher value of $p$ than $f_\chi(p)$, which brings its bulk distribution closer to (or even on) the cross-section resonance. 
\item A regime with $60.5\,{\rm GeV}\lesssim m_S\lesssim62.5$\,GeV that we will refer to 
as {\it resonant}. Here, we have $\langle \sigma v\rangle>\langle \sigma v\rangle_2$ 
around the time when the DM particles start to leave thermal equilibrium, because the larger 
mass combines with the relevant  momenta  to $s\sim m_h^2$. At slightly later times, on the other hand, still 
relevant in changing the DM abundance, the DM momenta have redshifted so much that we 
are back to a situation where typically $\sqrt{s}\lesssim m_h -\Gamma_h$ and 
hence $\langle \sigma v\rangle<\langle \sigma v\rangle_2$.  
\item Finally, there is a {\it super-resonant regime} with $62.5\,{\rm GeV}\lesssim m_S\lesssim65$\,GeV,
where decoupling occurs at such high temperatures that we have 
$\langle \sigma v\rangle>\langle \sigma v\rangle_2$ during the whole time it takes for $Y(x)$ to 
reach its asymptotic value (determining the relic density).
\end{enumerate}

\begin{figure*}
  \includegraphics[trim={0.5cm 1cm 1cm 1cm}, width=0.65\columnwidth]{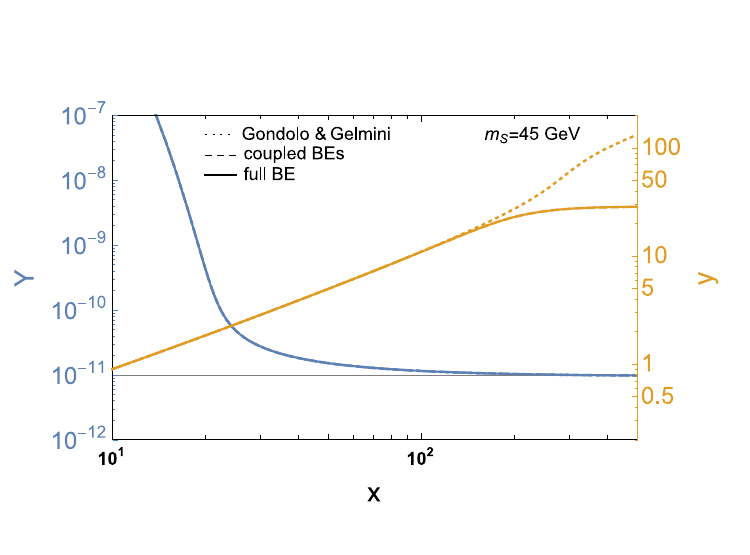}
  \hspace{0.3cm}
  \includegraphics[trim={0.5cm 1cm 1cm 1cm}, width=0.65\columnwidth]{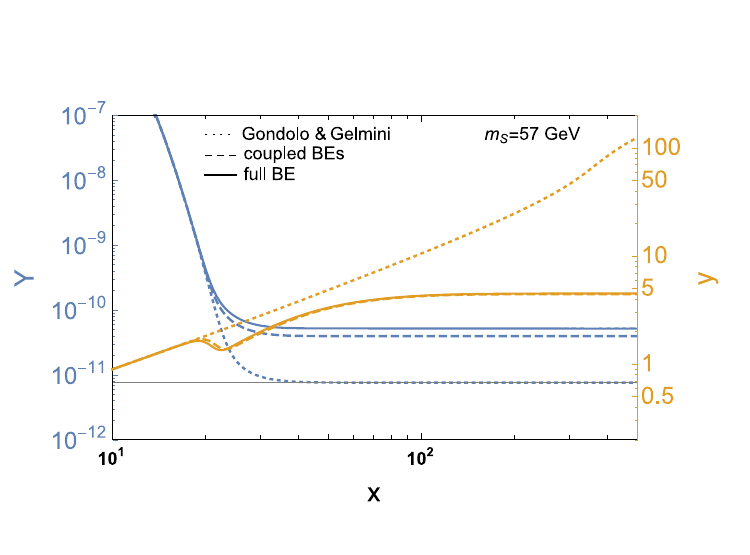}
  \hspace{0.3cm}
  \includegraphics[trim={0.5cm 1cm 1cm 1cm}, width=0.65\columnwidth]{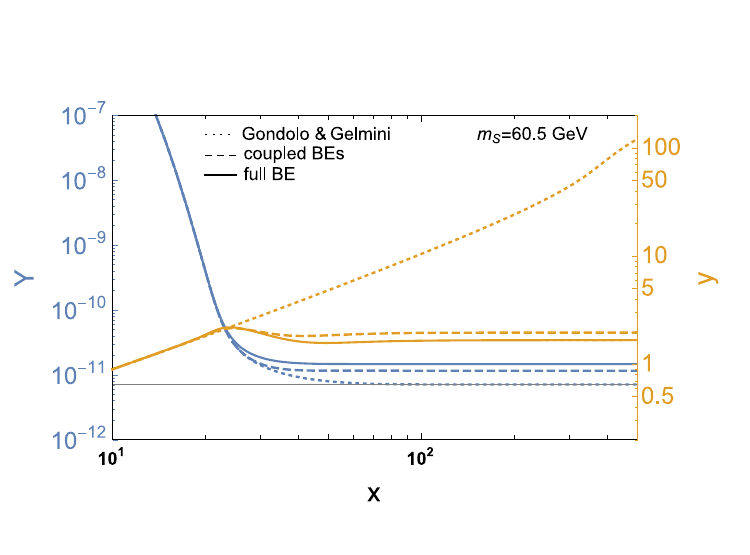}\\
  \includegraphics[trim={0.5cm 1cm 1cm 1cm}, width=0.65\columnwidth]{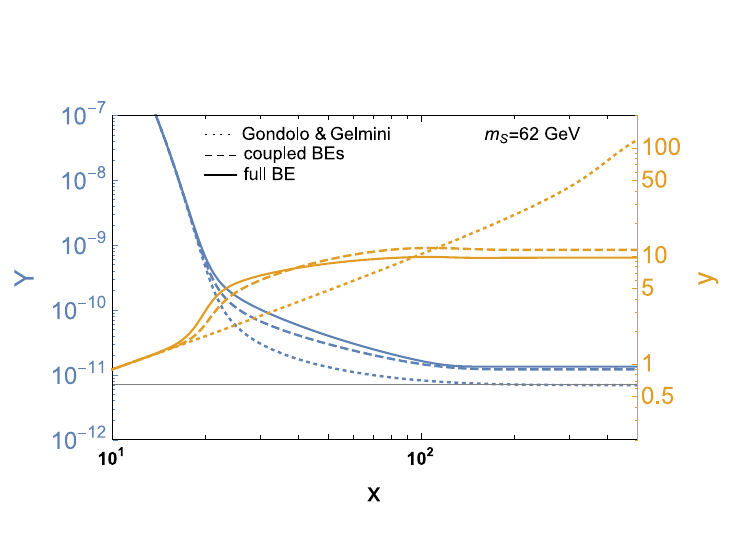}
  \hspace{0.3cm}
  \includegraphics[trim={0.5cm 1cm 1cm 1cm}, width=0.65\columnwidth]{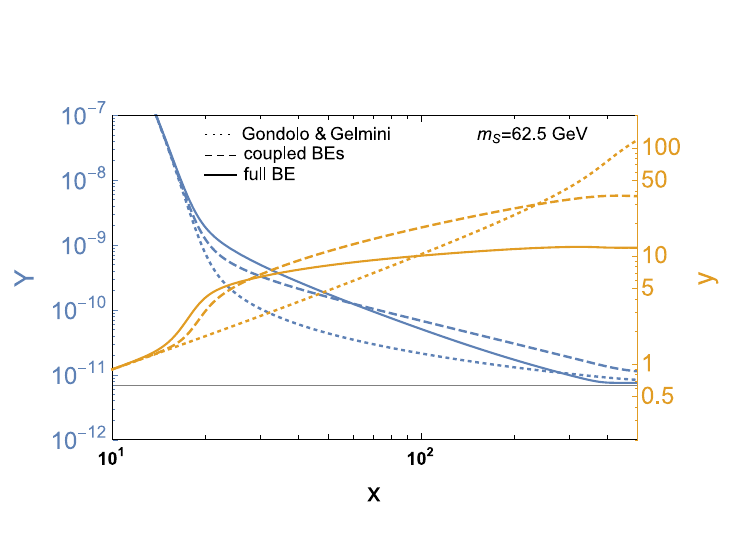}
  \hspace{0.3cm}
  \includegraphics[trim={0.5cm 1cm 1cm 1cm}, width=0.65\columnwidth]{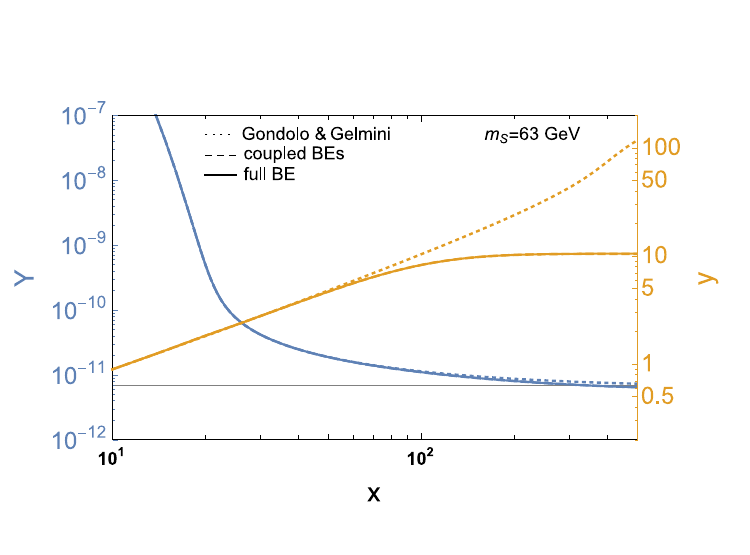}
  \caption{Evolution of $Y(x)$ and $y(x)$, for the same DM masses as shown in Fig.~\ref{fig:PSdist_multi}.
}
    \label{fig:yYx} \vspace{-0.3cm}
\end{figure*}

To help our discussion, let us look at a selection of benchmark points with Scalar Singlet masses
$m_S=45,57,60.5, 62,62.5,63$\,GeV and coupling constants $\lambda_S(m_S)$ that 
result in the correct relic density in the standard approach (dotted line in Fig.~\ref{fig:RDcont1}).
In Fig.~\ref{fig:PSdist_multi}, 
we show the DM distribution function for these benchmark points
that we find with our full numerical  approach, for selected values of $x$, and in Fig.~\ref{fig:yYx} 
the full evolution of $Y(x)$ and $y(x)$ for the different approaches. These figures thus extend the information in Fig.~\ref{fig:PSdist} by covering a range of DM masses.

The first thing to note, as exemplified by the benchmark points with $m_S=45$\,GeV and 
$m_S=63$\,GeV, is that for masses sufficiently far away from the resonance we find a 
phase-space distribution which remains almost exactly Maxwellian in shape. 
For these points, we therefore find as expected a very good agreement 
for the evolution of $Y(x)$ and $y(x)$ when comparing the numerical solution and the 
coupled Boltzmann approach, as well as with $Y$ in the standard Gondolo \& Gelmini setup 
(which assumes $T=T_\chi$). 
We note that this provides an important consistency check for both methods.

An example for a model in the {\it sub-resonant} region is the case with $m_S=57$\,GeV, which we
discussed in the main text. Here, the resonant annihilation depletes  $f_\chi(q)$ 
for momenta {\it just above the peak} of the distribution, leading to a relative decrease with respect
to a thermal distribution at these momenta, and hence a decrease in the DM velocity
dispersion (aka `temperature'). This effect is visible in Fig.~\ref{fig:PSdist_multi} starting with
a slight suppression at $q\sim8$ for the curve with $x=20$ 
(note that the relative enhancement 
at larger values of $q$ is not relevant  for our discussion given that $f_\chi$ is already highly 
suppressed here), and results in the decrease in the evolution of $y$ seen in Fig.~\ref{fig:yYx}. The latter
can also directly be understood from inspection of Eq.~(\ref{yfinalfinal}): in the sub-resonant 
regime we have $\langle \sigma v\rangle<\langle \sigma v\rangle_2$, which drives $y$ to smaller
values after decoupling (with a strength proportional to $Y$ -- which explains why the scattering term 
$\propto\gamma(T)$ can increase $y$ again, slightly,  once the DM abundance has decreased sufficiently).
A second effect of this depletion in $f_\chi(q)$ 
is that $\langle \sigma v\rangle_\mathrm{neq}$ 
decreases, which in turn leads to an earlier chemical decoupling and hence an increased relic 
density. 
The difference between the numerical and the coupled Boltzmann approach 
can in this case thus exclusively be understood as resulting from the slight offset in the 
$y(x)$ curves during the freeze-out
(which in turn results from the fact that the scattering term is not strong enough to maintain 
an exact Maxwellian shape of $f_\chi(q)$  when the velocity dispersion decreases as explained above.)

As we increase the DM mass, we leave the sub-resonant regime and enter the {\it resonant regime}, 
with the transition point marked by the benchmark model with $m_S=60.5$\,GeV. We note that this 
transition is also clearly visible in Fig.~\ref{fig:TkdTcd}, as a sharp decrease in the temperature
at which the DM velocity dispersion deviates from its equilibrium value.
The origin of this feature is {\it not} an actual delay of kinetic decoupling, but that DM annihilation 
now starts to deplete $f_\chi(q)$ {\it below} the peak of the would-be Maxwellian distribution.\footnote{
In a similar way, the sharp rise around $m_S\sim 54$\,GeV in Fig.~\ref{fig:TkdTcd} 
should not be interpreted as a feature in the momentum exchange rate $\gamma(T)$.
Rather, it can be understood as the point where the shape of the $y(x)$ evolution starts to develop 
from something close to the one in the top left panel in Fig.~\ref{fig:yYx} into something that is 
much closer to the one in the top center panel (which in turn is driven by the annihilation terms, as
explained in the text). As a result, the temperature at which $y$ departs 
from $y_\mathrm{eq}$ increases very quickly as the mass increases beyond this
transition point.
}
This leads to an increase of the velocity dispersion, once equilibrium is left, rather than a decrease as in the 
sub-resonant regime. This effect is very clearly seen in Figs.~\ref{fig:PSdist_multi} and \ref{fig:yYx}, 
up to DM masses at the 
higher end of this regime, where the influence of the resonance starts to become less important
because we have $\sqrt{s}\lesssim m_h+\Gamma_h$ only for DM momenta well below the peak of 
the phase-space distribution. 

In the {\it super-resonant regime} with $m_S\gtrsim m_h/2$, finally, we have necessarily
$\sqrt{s}\gtrsim m_h$. A resonantly enhanced annihilation rate  is thus only possible for a 
very small portion of phase-space, with almost vanishing relative DM momenta. This implies not
only that we always have $\langle \sigma v\rangle> \langle \sigma v\rangle_2$ in this regime, 
but also that the effect of the resonance rapidly becomes negligible.

Lastly, it is interesting to note that for $\sqrt{s}\gtrsim m_h$ the annihilation rate effectively  
features a  $1/v^2$ velocity dependence. This is similar to resonant Sommerfeld-enhanced annihilation, 
which leads to a suppressed relic density after a prolonged freeze-out phase \cite{vandenAarssen:2012ag}.
This can clearly be seen in the evolution of $Y(x)$ in Fig.~\ref{fig:yYx}, for $m_S\sim m_h/2$, 
where the differences between the numerical and
the coupled Boltzmann approach are mostly due to the late-time differences in $y(x)$ -- which in 
turn come about because of the rather significant differences in $f_\chi(q)$ at large values of $x$
(c.f.~Fig.~\ref{fig:PSdist_multi}).

\section{Semi-relativistic kinetic theory}
\label{app:semirel}

In this Appendix, we discuss how to generalize the highly non-relativistic elastic scattering term in Eq.~(\ref{Cresult})
to incorporate the most important relativistic corrections needed for the numerical implementation of 
the full Boltzmann equation. 
Throughout, we refer to this result as `semi-relativistic' scattering.

The starting point is to expand the full collision term $C_{\text{el}}$ in small momentum 
transfer compared to the typical DM momentum -- similar to what is done in order to arrive
at Eq.~(\ref{Cresult}), but not only keeping lowest-order terms in $\mathbf{p}^2/m_\chi^2\sim T/m_\chi$. 
From this, we can derive
a Fokker-Planck scattering operator in a relativistic form (for details, see \cite{Binder:2016pnr}):
\be
C_\mathrm{el}\simeq
\frac{E}{2} \nabla_{\mathbf{p}} \cdot
{\Bigg [}
\gamma(T,\mathbf{p})
\left (
E T  \nabla_{\mathbf{p}}  + \mathbf{p}
\right ) f_{\chi}
{\Bigg ]}
\,.
\label{eq:fokkerplanck1}
\ee
Being a total divergence, this scattering operator manifestly respects number conservation, as it should.
Another important property, which one can directly read off from the part inside the brackets, is that it
features a stationary point given by the relativistic Maxwell-Boltzmann distribution,
\be
\label{expET}
f_{\chi}^{\text{eq}} \propto e^{- E/T}.
\ee
The non-relativistic limit of Eq.~(\ref{eq:fokkerplanck1}) gives the scattering operator (\ref{Cresult}), but in this limit the 
stationary point would instead be the non-relativistic version 
 $f_{\chi}^{\text{eq}} \propto \exp[-p^2/(2m_\chi T)]$  --- 
which would cause a problem in the full BE as this does not correspond to the {\it actual} equilibrium distribution fed 
into the annihilation term of  Eq.~\eqref{eq:BEps}.

In general,  the momentum transfer rate $\gamma(T,\mathbf{p})$ in Eq.~(\ref{eq:fokkerplanck1}) 
depends on the DM momentum $\mathbf{p}$. However, the stationary point is independent of 
$\gamma$, which motivates us to restrict ourselves to the leading order term 
$\gamma(T) \equiv \gamma(T,\mathbf{0})$, neglecting any momentum dependence, and 
use the non-relativistic limit in Eq.~(\ref{eq:fokkerplanck1}) \emph{only} to evaluate the momentum transfer rate 
$\gamma(T)$ as it appears in Eq.~(\ref{cTdef}). To this order, we could thus also replace the leading $E$ in 
Eq.~(\ref{eq:fokkerplanck1}) by $m_\chi$; here, we choose to still keep it as it leads to a much more compact 
analytical form of the equation governing the DM temperature (see below). 
Explicitly performing the first partial derivative in $C_{\text{el}}$ then leads to the final form of our 
semi-relativistic Fokker-Planck operator as given by Eq.~(\ref{Csemirel}). 
This operator is our default choice for the numerical implementation of the full Boltzmann equation.

As already pointed out in Section \ref{sec:magicmichael}, it is mandatory for the full phase-space 
calculation to have a scattering operator with a fixpoint that matches the equilibrium distribution of
Eq.~(\ref{expET}) assumed in the annihilation term. For the coupled integrated Boltzmann system, 
on the other hand, this issue is fully addressed by using the relativistic temperature definition of 
Eq.~(\ref{tdef}) --- rather than its non relativistic version typically adopted in the literature in the 
context of kinetic decoupling --- because this automatically leads to the correct fixpoint $T_\chi=T$ 
for both the semi-relativistic Eq.~(\ref{Csemirel}) and, to the lowest order, for the non-relativistic 
version Eq.~(\ref{Cresult}); see the discussion in Section \ref{sec:cBE}.

Another advantage of our semi-relativistic Fokker-Planck operator is that the differential equation for 
$T_\chi$, often quoted when discussing kinetic decoupling, takes a very simple form even 
beyond the highly non-relativistic limit.
To see this, let us for the moment ignore the impact of annihilations, and take the second moment of
the Boltzmann equation with this operator (using the relativistic definition of $T_\chi$).
This leads to 
\begin{align}
& \dot{T}_{\chi} + 2 \left(1- \frac{\langle p^4/E^3 \rangle}{6 T_{\chi}}\right) H T_{\chi}  = \\ 
& \gamma {\Bigg [} T \left(1- \frac{5}{6}\langle p^2/E^2\rangle + \frac{2}{6}\langle p^4/E^4\rangle \right)  - T_{\chi} \left(1- \frac{\langle p^4/E^3 \rangle}{6 T_{\chi}}\right){\Bigg ]}\,,\nonumber
\end{align}
which of course is equivalent to Eq.~(\ref{yfinalfinal}) in the main text, when neglecting the annihilation 
terms and implementing the replacement given in Eq.~(\ref{gamma_semirel}).
Let us repeat that the r.h.s.~of the above equation only takes this particular form with our default 
choice of the semi-relativistic Fokker-Planck 
term, whereas the moment appearing on the left hand side is an exact result. This equation is in 
general not closed in terms of $T_\chi$. 
However, if we make the ansatz of a Maxwellian DM phase-space distribution, 
c.f.~Eq.~(\ref{MB_ansatz}),  we get a relation between the different momentum moments,
\begin{align}
5 \langle p^2/E^2\rangle - 2 \langle p^4/E^4\rangle = {\langle p^4/E^3 \rangle}/{T_{\chi}},
\end{align}
such that the differential equation closes in terms of $T_{\chi}$. Indeed, introducing
\be
2(1-w)\equiv \frac{g_\chi}{3T_\chi n_\chi}\int \frac{d^3p}{(2\pi)^3}\,\frac{\mathbf{p}^4}{E^3} f_\chi(\mathbf{p})
= \frac{\langle p^4/E^3 \rangle}{3T_\chi} \,,\label{wdef}
\ee
it takes a very simple form:
\begin{align}
\dot{T}_{\chi} +  2 w(T_{\chi}) H T_{\chi}  =  w(T_{\chi}) \gamma(T) \left( T - T_{\chi}\right).\label{eq:semireltemperature}
\end{align}
This generalizes the highly nonrelativistic result  \cite{Bringmann:2006mu}, for which $w\to1$ and we hence find 
the familiar scaling $T_{\chi} \propto T^2$ after kinetic decoupling (i.e.~when $\gamma \ll H$). In the
ultra-relativistic limit, on the other hand, we have $w(T_{\chi}) \rightarrow 1/2$ and the likewise familiar
scaling of $T_{\chi} \propto T$ for relativistic particles. We note that in the region $x \gtrsim 10$ 
relevant for early kinetic decoupling, the correction to the non-relativistic limit is already sizeable; 
e.g.~$w(x=10) \approx 0.8$.

\bibliography{EarlyKD}

\end{document}